\documentclass[fleqn,usenatbib]{mnras}

\usepackage{newtxtext,newtxmath}
\usepackage{mathptmx}

\usepackage[T1]{fontenc}
\usepackage{ae,aecompl}

\usepackage{graphicx}	
\usepackage{amsmath}	
\usepackage{amssymb}	

\usepackage{subfigure}
\usepackage{cancel}
\usepackage[title]{appendix}

\title[Tidal disruptions of planetary bodies I]{Tidal disruption of planetary bodies by white dwarfs I: A hybrid SPH-analytical approach}

\author[Malamud \& Perets]
{		
	Uri Malamud$^{1,2}$ and
	Hagai B. Perets$^{1,3}$
	\\	
	$^{1}$Department of Physics, Technion - Israel Institute of Technology, Technion City, 3200003 Haifa, Israel\\
	$^{2}$School of the Environment and Earth Sciences, Tel Aviv University, Ramat Aviv, 6997801 Tel Aviv, Israel\\
	$^{3}$TAPIR, California Institute of Technology, Pasadena, CA 91125, USA
}

\date{Accepted XXX. Received YYY; in original form ZZZ}

\pubyear{2019}

\begin{document}
	
\label{firstpage}
\pagerange{\pageref{firstpage}\textendash{}\pageref{lastpage}}
\maketitle
	
\begin{abstract}
We introduce a new hybrid method to perform high-resolution tidal disruption simulations, at arbitrary orbits. An SPH code is used to simulate tidal disruptions only in the immediate spatial domain of the star, namely, where the tidal forces dominate over gravity, and then during the fragmentation phase in which the emerging tidal stream may collapse under its own gravity to form fragments. Following each hydrodynamical simulation, an analytical treatment is then applied to instantaneously transfer each fragment back to the tidal sphere for its subsequent disruption, in an iterative process. We validate the hybrid model by comparing it to both an analytical impulse approximation model of single tidal disruptions, as well as full-scale SPH simulations spanning the entire disc formation. The hybrid simulations are essentially indistinguishable from the full-scale SPH simulations, while computationally outperforming their counterparts by orders of magnitude. Thereby our new hybrid approach uniquely enables us to follow the long-term formation and continuous tidal disruption of the planet/planetesimal debris, without the resolution and orbital configuration limitation of previous studies. In addition, we describe a variety of future directions and applications for our hybrid model, which is in principle applicable to any star, not merely white dwarfs.
\end{abstract}
	
\begin{keywords}
	planets and satellites: terrestrial planets, hydrodynamics, stars: white dwarfs
\end{keywords}

\section{Introduction}\label{S:Intro}
Given the short sinking timescale of elements heavier than helium in the atmospheres of WDs \citep{Koester-2009}, the large fraction of WDs that are polluted with heavy elements \citep{ZuckermanEtAl-2003,ZuckermanEtAl-2010,KoesterEtAl-2014} is readily explained by accretion of planetary material \citep{DebesSigurdsson-2002,Jura-2003,KilicEtAl-2006,Jura-2008}. The current view, based on the inferred composition of both WD atmospheres \citep{WolffEtAl-2002,DufourEtAl-2007,DesharnaisEtAl-2008,KleinEtAl-2010,GansickeEtAl-2012,JuraYoung-2014,HarrisonEtAl-2018,HollandsEtAl-2018,DoyleEtAl-2019, SwanEtAl-2019} and their discs \citep{ReachEtAl-2005,JuraEtAl-2007,ReachEtAl-2009,JuraEtAl-2009,BergforsEtAl-2014,Farihi-2016,ManserEtAl-2016,DennihyEtAl-2018} suggests that the polluting material is terrestrial-like and typically dry.
	
Orbiting dust is deduced from measurements of infrared excess, while gas is inferred from metal emission lines. The spatial distribution of the gas is typically within the WD tidal disruption radius, and it often orbits the star with some eccentricity \citep{GansickeEtAl-2006,GansickeEtAl-2008,DennihyEtAl-2016,DennihyEtAl-2018,CauleyEtAl-2018}. The origin of material at such close proximity to the WD is clearly not primordial \citep{GrahamEtAl-1990}, since the WD disruption radius is of the order of the progenitor star's main-sequence physical radius \citep{BearSoker-2013}. It is instead thought to originate from planetary bodies which are perturbed by some mechanism \citep{DebesSigurdsson-2002,BonsorEtAl-2011,DebesEtAl-2012,KratterPerets-2012,PeretsKratter-2012,ShapeeThompson-2013,MichaelyPerets-2014,VerasGansicke-2015,StoneEtAl-2015,HamersPortegiesZwart-2016,Veras-2016,PayneEtAl-2016,CaiazzoHeyl-2017,PayneEtAl-2017,PetrovichMunoz-2017,StephanEtAl-2017,SmallwoodEtAl-2018} to highly eccentric orbits with proximity to the WD, and are subsequently tidally disrupted to form a circumstellar disc of planetary debris.
	

To date, there exist very few detailed simulations of disc formation by tidal disruptions. The study of \cite{VerasEtAl-2014} constitutes the most detailed and relevant work thus far, which investigates the initial formation of white dwarf debris discs, caused by the tidal disruption of kilometer-sized asteroids ($\sim10^{14}$ kg). It follows a similar study by \cite{DebesEtAl-2012} which only considered the first initial tidal disruption of an extremely eccentric asteroid instead of the entire debris disc formation, while both studies used the same modified N-body code (\emph{PKDGRAV}). Under the conditions discussed in the \cite{VerasEtAl-2014} paper, the disrupted asteroid debris fill out a highly eccentric ring of debris, along the original asteroid trajectory. The material in this disc does not immediately accrete onto the white dwarf at this early stage, and instead the disc is required to evolve further, perhaps through various radiation processes \citep{VerasEtAl-2015}, into a more compact state. 

Being conceptually similar, the study of \cite{WeissmanEtAl-2012} (based on the N-body code developed by \cite{MovshovitzEtAl-2012}) investigates the tidal disruption of the Sun-grazing, Kreutz-family progenitor. Their results show that the disruption around the Sun breaks the object into multiple clumps, depending on the exact density and perihelion distance assumed. They suggest that the observed size distribution of the Kreutz group can perhaps be produced, however multiple returns are needed by the parent object and its initial ensuing fragments in order to provide the observed temporal separation of major fragments. The \cite{WeissmanEtAl-2012} and \cite{DebesEtAl-2012} studies underline the difficulties of modelling such tidal disruptions -- since these objects are often highly eccentric (with $e$ approaching 1), time-step limitations make tracking of the entire orbit for multiple returns computationally implausible. One can substantially reduce either the resolution or the eccentricity (or both) in order to circumnavigate this problem, which was precisely the solution adopted by \cite{VerasEtAl-2014} in order to enable multiple returns (orbits) in their simulations.

At the opposite end of the planetary size distribution, some studies consider the tidal disruption of gas giants, demonstrating the exact same problem. \cite{FaberEtAl-2005,LiuEtAl-2013} use an SPH code in order to simulate a close gas giant flyby around a star (i.e. a single tidal encounter) whereas \cite{GuillochonEtAl-2013} use a grid-based code and considered both single as well as multiple passage encounters. As in the \cite{VerasEtAl-2014} study, multiple returns were accomplished only by considerably lowering the assumed eccentricity of the planets.

Using a very simple analytical model, we demonstrate in Section \ref{S:Analytical} that one cannot simply change important characteristics like the eccentricity or size of the disrupted progenitor without directly (and substantially) affecting the properties of the debris that are produced by the tidal disruption. We therefore emphasize the main shortcomings of all previous studies:

\begin{itemize}
	\item no previous study has investigated the detailed disruption of terrestrial-sized or dwarf-sized planets, despite being potentially important in terms of the typically inferred composition of the pollutants, the effect of having larger-than-asteroid size on the outcome of the disruption, the recent determination of oxygen fugacities which suggest that polluting rocky materials are geophysically and geochemically similar to Earth \citep{DoyleEtAl-2019} and the implications of what could be a stripped core from a larger original object \cite{ManserEtAl-2019}.
	\item the resolution in previous studies is orders of magnitude lower (few $\sim 10^3$ particles) than the standard resolution currently used in modern SPH or N-body applications ($\sim 10^5-10^6$), due to the aforementioned time-step limitation.
	\item in order to enable multiple returns the orbital parameters of the disrupted parent-bodies are contrived. Reducing the semi-major axis and eccentricity changes the outcome of tidal disruptions and in turn the ensuing debris discs.
	\item studies that alternatively did consider realistic orbits, were instead limited to calculating only the first tidal encounter, whereas the full disc formation typically requires multiple returns.	
\end{itemize}

The goal of this study is therefore to resolve such difficulties by utilizing a new, hybrid concept, to modelling tidal disruptions. Our approach is to omit unnecessary calculations far from the vicinity of the star, by fully following the disruption and coagulation of particles into fragments with SPH, \emph{only} when they are within the star's immediate environment. For the reminder of their orbits, fragment trajectories are calculated and tracked analytically assuming keplerian orbits. I.e., we make the assumption (for simplicity) that the disc of debris is largely collisionless, as well as dynamically unaffected by radiation or other processes, and then instantaneously transfer the fragments back to the tidal sphere for their next flyby, in an iterative process. Our assumptions are discussed and quantified. 

With each disruption, the semi-major-axis dispersion of newly formed fragments depends on the exact size and orbit of their progenitor. The hybrid code handles the synchronization, timing and dissemination of SPH jobs. The disc formation completes only when reaching one of two outcomes: either all fragments have ceased disrupting given their exact size, composition and orbit; or fragment disruption is inhibited when reaching the numerical minimum size - that of a single SPH particle. 

The hybrid approach enables studying tidal disruptions for any progenitor orbit (even objects originating from tens or hundreds of AU) with the same efficiency. The code easily handles partial disruptions (i.e., those resulting in little mass shedding when the pericentre distance is sufficiently large), since now the number of iterations is not limited by a large semi-major axis.

The layout of the paper is structured as follows: in Section \ref{S:Analytical} we first outline an analytical model of tidal disruptions. This model provides invaluable insights about the outcome of single tidal disruptions. While discussing its limitations, we develop a deeper understanding of what may be expected in performing full numerical tidal disruption simulations involving terrestrial-sized or dwarf-sized planets; In Section \ref{S:FullSPH} we then perform full-scale tidal disruption simulations using SPH. We discuss the code details and setup, show the various disruption outcomes which depend on our choice of pericentre distance, track the formation of the disc and examine the effect of applying initial rotation to the disrupted planet; In Section \ref{S:Hybrid} we introduce our hybrid model, describing its principles and the validity of its assumptions. We then verify and corroborate our hybrid model results against full SPH simulations, showing that the two methods are in agreement, while discussing how the hybrid method outperforms the former. We show that unlike previous tidal disruption studies of small asteroids which form ring-like structures on the original orbit, larger bodies form dispersed structures of interlaced elliptic eccentric annuli on tighter orbits. In Section \ref{S:Future} we discuss various  different applications and future improvements for our hybrid model. Finally, in Section \ref{S:Summary} we summarize the paper's main achievements. In an accompanying paper (Malamud and Perets 2019; hereafter Paper II) we utilize the new code as to consider a suite of simulations of the tidal disruptions of rocky bodies by WDs, spanning a large range of masses, semi-major axes and pericentre distances, analyze them and discuss the results.

\section{Analytical impulsive disruption approximation}
\label{S:Analytical}
The tidal disruption of a planetesimal can be approximated analytically via an impulsive disruption. It entails the assumptions that (1) a spherically symmetric planetesimal remains undisturbed until it reaches the distance of closest tidal approach; (2) it then instantaneously breaks into its constituent particles; (3) the latter retain their previous centre-of-mass velocity, albeit now occupy a range of spatial coordinates; and (4) it is assumed that the constituent particles evolve independently of each other immediately after the breakup, tracing out ballistic trajectories in the star's gravitational potential.

The impulsive disruption approximation provides a rather simple analytical framework for gaining insight and intuition regarding the fundamental disruption properties, however, with the caveat that tidal breakup is never strictly and completely impulsive. As we will show, our assumptions break down considerably, depending primarily on the distance of close tidal approach.

In what follows, consider a central star of mass $M$, orbited by a planetesimal of mass $m$ which undergoes an impulsive tidal disruption at distance $d$ from the star. At the moment of breakup, the planetesimal has the velocity $v$ and semi-major axis $a$. The velocity $v$ is given by the 'vis-viva' equation, for any keplerian orbit, in the form:

\begin{equation}
v = \sqrt{\mu\left( \frac{2}{d}-\frac{1}{a} \right)}
\label{eq:VisVisa}
\end{equation}	

where $\mu=G(M+m)$ is the standard gravitational parameter, and $G$ the gravitational constant. Since we assume that an arbitrary particle's velocity $\acute{v}$ equals the previous centre-of-mass velocity of the whole planetesimal $v$ at the moment of breakup, neglecting the velocity of self-rotation of the planetesimal (typically two orders of magnitude lower than the centre-of-mass velocity even for a planet sized object), the two can be equated such that:

\begin{equation}
G(M+\acute{m})\left( \frac{2}{d+r}-\frac{1}{\acute{a}} \right) = G(M+m)\left( \frac{2}{d}-\frac{1}{a} \right)
\label{eq:EquateVel}
\end{equation}

where $\acute{m}$ is the particle mass, $\acute{a}$ its semi-major axis and $r$ its displacement relative to the planetesimal's centre-of-mass at breakup, such that $\acute{d}=d+r$. We assume a spherically symmetric planetesimal, hence the maximal displacement equals the planetesimal's radius $R$, such that $\lvert r \rvert \leq R$.

Let us assume that $\acute{m}\ll m$ and $m\ll M$. The latter assumption is highly judicious given the typical mass of terrestrial planets or less. Hence equation \ref{eq:EquateVel} can be re-written to extract the particle's semi-major axis as a function of its displacement $r$:

\begin{equation}
\acute{a} = a  \left( 1-a\frac{2r}{d(d+r)} \right)^{-1}
\label{eq:SemiMajorAxis}
\end{equation}

When the denominator equals zero, particles assume a parabolic trajectory. The critical displacement $r_\mathrm{crit}$ for which it occurs, equals:

\begin{equation}
r_\mathrm{crit} = \frac{d^2}{2a-d}
\label{eq:Rcrit}
\end{equation}

\begin{figure}
    \begin{center}	
        \includegraphics[scale=0.5]{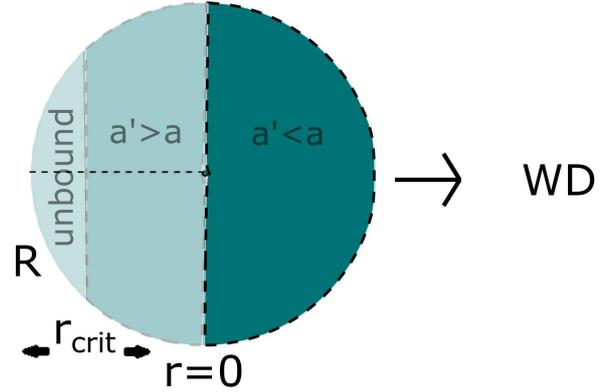}
		\caption{The disruption roadmap: the semi-major axes of disrupted constituent particles within a spherically-symmetric planetesimal are determined according to their various locations within the object. The exact delineation of $r_\mathrm{crit}$ is pivotal, and determines different disruption outcomes. As indicated by the short arrows it may move in either direction. The right hemisphere (black dashed line) is always bound. If $0<r_\mathrm{crit}<<R$, the left hemisphere is entirely unbound, whereas if $r_\mathrm{crit}>R$ both hemispheres are bound, corresponding to the regimes outlined in Table \ref{tab:regimes}}
	 	\label{fig:roadmap}
	\end{center}	
\end{figure}

Particles with $r>r_\mathrm{crit}$, i.e. particles which are sufficiently displaced from the center-of-mass in the opposite direction of the WD, will become unbound. Particles with exactly $r=0$ will satisfy $\acute{a}=a$, keeping the original semi-major-axis. Particles with $0<r<r_\mathrm{crit}$ will have larger-than $a$ semi-major axes, and all particles with $r<0$ (in the direction of the WD with respect to the centre-of-mass) necessarily have $\acute{a}<a$. The disruption 'roadmap' is visually presented in Figure \ref{fig:roadmap}, depicting the different parts of a cross-section, of a disrupted spherical planetesimal.
 
\begin{figure}
	\begin{center}	
		\includegraphics[scale=0.558]{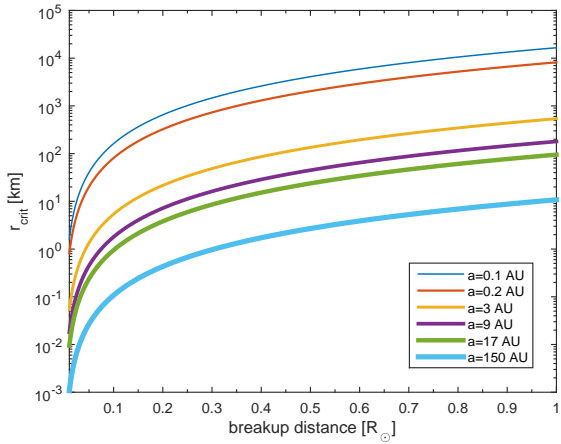}
		\caption{The critical displacement for generating unbound particles, shown in logarithmic scale as a function of the breakup distance from the star. Six lines denoted by different colours and widths correspond to various $a$ (semi-major axes) of the original planetesimal. Large semi-major axes are more likely to result in unbound disrupted material.}
		\label{fig:Rcrit}
	\end{center}	
\end{figure}

Let us further examine the critical displacement $r_\mathrm{crit}$. Since it is proportional to the square of the planetesimal breakup distance $d$, it can vary by several orders of magnitude. Hence, in Figure \ref{fig:Rcrit} we show $r_\mathrm{crit}$ as a function of $d$ in logarithmic scale, the latter ranging between the typical WD radius ($\sim 10^{-2}R_\odot$) to the typical WD Roche radius ($\sim R_\odot$). The plot features six lines with varying semi-major axes $a$ of the original planetesimal before breakup, spanning three orders of magnitude (between 0.1 AU and 150 AU), and corresponding to all the $a$ values simulated throughout this paper and Paper II. Figure \ref{fig:roadmap} is essential to determining the outcome of single tidal disruptions. It emphasizes the importance of the perturbing mechanisms that inject planetesimals to tidal crossing orbits, since both the breakup distance and the semi-major axis are the decisive dynamical factors that shape the outcomes of tidal disruptions.

Clearly there exist three distinct disruption regimes. If the parameters of impulsive disruptions are such that $r_\mathrm{crit}\ll R$, roughly half of the debris will be unbound from the system, while the other half will be placed onto much tighter orbits compared to the original $a$. The latter is easily seen, since $r_\mathrm{crit}\ll R$ can be re-written as:

\begin{equation}
\frac{2Ra}{d(d+R)} \gg  1
\label{eq:FirstRegieme}
\end{equation}

Using the condition from Equation \ref{eq:FirstRegieme} and applying it to Equation \ref{eq:SemiMajorAxis}, $\acute{a}$ assumes large negative values (hyperbolic trajectories) for positive displacement ($r>0$). For negative $r$ the original semi-major axis $a$ is divided by a large denominator ($\gg  1$), hence the tight orbits.

Moreover, note that Equation \ref{eq:SemiMajorAxis} can be re-written as $\acute{a}=\mp d(d+r)/2r$, such that $\acute{a}$ is independent of $a$, and particles converge onto a minimum semi-major axis value of $\acute{a}=\mp d^2/2R$ (since typically $R\ll d$). Such extremely \emph{bi-modal disruption} regimes are often formulated in many studies from an energy dispersion point of view (e.g. see \cite{MetzgerEtAl-2017}), where the particle's energy spread is 'frozen-in' at the moment of breakup. We note that the freezing point, or breakup distance $d$, is not necessarily interchangeable with the planetesimal's pericentre distance $q$, since the breakup does not necessarily occur at $q$ (although often studies indeed make that assumption). Several authors have previously demonstrated the importance of this point for stellar disruptions \citep{StoneEtAl-2013,GuillochonRamirez-Ruiz-2013,SteinbergEtAl-2019}. We likewise show in Paper II that $d$ may be different from $q$ for the kind of planetary disruptions considered in this study. Nevertheless, henceforward we sometimes do use $q$ and $d$ interchangeably, but only when referring to literature which does not make this distinction specifically.

The other extreme regime which satisfies $r_\mathrm{crit}\gg R$, necessarily places all the constituent particles in bound orbits. Furthermore, the dispersion in $\acute{a}$ is negligible by the same argument since the denominator $\simeq 1$. The \emph{non-dispersive disruption} therefore results in the formation of an eccentric ring on the original orbit $a$, filled up by debris. See e.g. the \cite{VerasEtAl-2014} study, where the disrupted object is a very small asteroid ($R \simeq 3$ km) with a semi-major axis of 0.2 AU and $q$ between 0.135$R_\odot$ and 0.27$R_\odot$, which according to Figure \ref{fig:Rcrit} (and assuming $d=q$) leads to $r_\mathrm{crit}$ of a few $10^2$ km - some two orders of magnitude larger than $R$. We note however that their semi-major axis of 0.2 AU is intentionally small, due to the computational limitations that our paper attempts to circumnavigate by using the hybrid model. Figure \ref{fig:Rcrit} demonstrates that the same asteroid only at a larger, more realistic semi-major axis, will no longer be in the \emph{non-dispersive disruption} regime and therefore it will not form an eccentric ring, but rather a more dispersed disc. For example, at $a=4.77$ AU and $q=0.24-0.32~R_{\odot}$, the outcome in the \cite{DebesEtAl-2012} study would have been an annulus instead of a ring (were it computationally possible to carry out the formation to its completion), since in this case $r_\mathrm{crit}$ is larger than but similar to the asteroid's $R$. If instead the \cite{VerasEtAl-2014} asteroid originated from an analogue Kuiper belt, we might even get a \emph{bi-modal disruption}.

The third \emph{intermediate disruption} regime entails $r_\mathrm{crit}\simeq R$, which results in some dispersion of the original semi-major axis, depending on the exact parameters of the problem. The ensuing disc is therefore the least straight-forward to characterize. For a summary of the different disruption regimes see Table \ref{tab:regimes}.

\begin{table}
	\caption{Tidal disruption regimes.}
	\begin{tabular}{*{3}{l|}}	
		\hline
		{Regime} & {Condition} & {Outcome} \\
		\hline
		\emph{bi-modal disruption}		& $r_\mathrm{crit}\ll R$ 	& $\acute{a}(r\to R) = \pm d^2/2R$ \\
		\emph{intermediate disruption}  & $r_\mathrm{crit}\simeq R$& $\acute{a}$ from Equation \ref{eq:SemiMajorAxis}\\
		\emph{non-dispersive disruption}& $r_\mathrm{crit}\gg R$ 	& $\acute{a}\cong a$\\
		\hline
	\end{tabular}
	\label{tab:regimes}
	The disruption regime is determined by the characteristic value of $r_\mathrm{crit}$ from Equation \ref{eq:Rcrit}. The new semi-major axes $\acute{a}$ of disrupted particles ranges between having two symmetric peaks in the distribution, to having no dispersion at all (i.e., keeping the original semi-major axis $a$).
\end{table}

What outcomes might then we expect to find in our simulations? In this paper, most of the scenarios investigated can be traced within Figure \ref{fig:Rcrit} to have $r_\mathrm{crit}<R$, where $r_\mathrm{crit}$ and $R$ typically differ by about one order of magnitude. Such disruptions are therefore neither \emph{bi-modal}, nor \emph{non-dispersive}, which only emphasizes why detailed simulations are required. Broadly speaking, they are nevertheless closer to the \emph{bi-modal} regime, and thus we might expect the outcome of our disruptions to display, at least in part, some kind of resemblance to a bi-modal semi-major-axis distribution.

There are, however, additional complications. We must remember that unlike in the impulsive disruption approximation, real disruptions do not abide by our set of assumptions. Planetesimal breakup is neither instantaneous nor complete, and the assumption of sphericity is violated. We note that the disruption chiefly depends on the tidal force which breaks the planetesimal apart. For $R\ll d$, the tidal force per unit mass $F_T$ can be approximated by:

\begin{equation}
F_T=\frac{2GMR}{d^3}
\label{eq:TidalForce}
\end{equation}

Since this tidal force greatly depends on the breakup distance $d$, a complete disruption is more likely to occur when the object passes close to the star. Consider for example a very deep tidal disruption with $q=$0.1$R_\odot$ versus a moderate one near the Roche limit with $q=$1$R_\odot$ (see discussion on the Roche distance in Section \ref{SS:SPHOutline}). The former leads to a tidal force 1000 times greater (tentatively assuming $d=q$), while the opposing force of self-gravity remains the same, thus we can expect a huge difference in the outcomes of these two cases (see Section \ref{SS:PericentreDependence} for a quantitative perspective). A common outcome in our simulations, unlike in the impulsive disruption approximation, is a partial, rather than a full disruption.

Since the disruption proceeds gradually, as the planetesimal's motion carries it deeper into the tidal sphere before reaching its closest approach, the tidal process is not instantaneous by definition, and there is always some measure of tidal elongation prior to breakup. Hence, one might consider a more realistic spatial distribution of the constituent particles, as opposed to the simple spherical view described above. This would alter the actual dispersion of the particles.

An additional complication is that the simple impulsive approximation does not really capture the subtleties and nuances of inhomogeneous planetesimals, as clumps of particles following a disruption can consist of different materials and/or have complex internal structures that vary in density (and strength, but we will omit that discussion for the moment). Different materials thus have various Roche radii.

Also, as previously mentioned, the orbit is usually well defined in tidal disruption problems, so we have good knowledge of $q$. However, we do not have good knowledge of $d$, and the 'instantaneous breakup', such as we have defined it, may actually occur prior to the closest approach. Without any sophisticated treatment, we often equate $d$ with $q$ as a heuristic approach, allowing us to draw simplistic analytical approximations.

Finally, as shown in Section \ref{SS:DiscFormation}, the outcome of real disruptions is modified to some extant by the self-rotation of the planetesimal. This effect is not negligible, especially for rapid self-rotation. 

The picture that emerges in real tidal disruptions therefore involves some dispersion of the original planetesimal semi-major axis. Whether full or partial, the disrupted clumps returning for an additional tidal passage, now occupy a range of different sizes, compositions, self-rotation rates and semi-major axes, with only their pericentre distance unchanged. They will therefore potentially follow a different disruption regime during each subsequent flyby, making the problem too complex for any simple approximative analytical model, emphasizing the importance of numerical modelling.

\section{Full SPH disruption simulations}\label{S:FullSPH}

\subsection{Code outline and setup}\label{SS:SPHOutline}
We perform hydrodynamical disruption simulations using an SPH code developed by \citet{SchaferEtAl-2016}. The code is implemented via CUDA, and runs on graphics processing units (GPU), with a substantial improvement in computation time, on the order of several $\sim10^{1}-10^{2}$ times faster for a single GPU compared to a single CPU, depending on the precise GPU architecture. The code has already been successfully applied to several studies \citep{DvorakEtAl-2015,MaindlEtAl-2015,HaghighipourEtAl-2016,WandelEtAl-2017,BurgerEtAl-2018,HaghighipourEtAl-2018,MalamudEtAl-2018}.

The code implements a Barnes-Hut tree that allows for treatment of self-gravity, as well as gas, fluid, elastic, and plastic solid bodies, including a failure model for brittle materials and a treatment for small porous bodies. Here we perform our simulations while neglecting solid-body physics, being more computationally expensive. We however lay out future plans (Section \ref{SS:Strength}) to also perform a dedicated study including material strength, outlining its potential importance. We use the Tillotson equation of state (EOS). The parameters for the EOS are taken from \citet{Melosh-1989} and \citet{BenzAsphaug-1999}, for iron and silicate (basalt) respectively. See \cite{MalamudEtAl-2018} for further details.

Throughout this Section we will perform full hydrodynamical simulations of planetesimals which undergo tidal disruption around a 0.6 M$_{\odot}$ WD. The star mass is chosen to correspond to the peak mass in the observed WD mass distribution. It is a common practice in many WD studies to adopt this fiducial WD mass \citep{TremblayEtAl-2016,Veras-2016,CummingsEtAl-2018}. The disrupted planetesimal mass is treated as a free parameter, however in this section we only simulate the tidal disruption outcome of planets with masses corresponding to that of Mars and Earth, or 0.1 M$_{\oplus}$ and 1 M$_{\oplus}$ respectively. For simplicity we consider all planetesimals to have an Earth-like composition and structure, being differentiated and composed of 30\% iron and 70\% dunite by mass. 

As discussed in Section \ref{S:Analytical}, the outcome of a tidal disruption is highly dependent on its depth. That is, when the pericentre distance $q$ is a smaller fraction of the Roche limit, the event is more likely to break the object down to its constituent particles (as in the impulse approximation set of assumptions), while a more grazing passage will result in a partial disruption that breaks only the planetesimal's outer portions. In order to investigate and compare such differences we consider in Section \ref{SS:PericentreDependence} the following $q$ values: 0.1$R_\odot$, 0.5$R_\odot$ and 1$R_\odot$ (or 10\%, 50\% and 100\% of a Roche grazing orbit, respectively). The considerations for the pivotal 1$R_\odot$ grazing orbit are discussed below.

Throughout most of the paper we consider \emph{initially} non-rotating planetesimals (i.e., referring strictly to the initial rotation, but not the later rotation of tidally disrupted fragments). This assumption is however tested and evaluated in Appendix \ref{A:RotationDependence}, where we assume the planet to have a 20 h rotation period in both the prograde and retrograde senses prior to the disruption, and compare the outcomes to that of non-rotating planets.

Each simulation starts with a relaxed planetesimal internal structure, i.e. having hydrostatic density profile and internal energy values from adiabatic compression, following the algorithm provided in appendix A of \citet{BurgerEtAl-2018}. This self-consistent semi-analytical calculation (i.e., using the same constituent physical relations as in the SPH model) equivalently replaces the otherwise necessary and far slower process of simulating each body in isolation for several hours, letting its particles settle into a hydrostatically equilibrated state prior to the collision (as done e.g., in the work of \citet{CanupEtAl-2013} or \citet{SchaferEtAl-2016}). Since the relaxation algorithm does not account for the additional effect of rotation, in Appendix \ref{A:RotationDependence} we place initially rotating planetesimals far from the star, providing them with an extra $\sim$30 h relaxation phase that damps any residual radial oscillations before the planet approaches the star.

For all other simulations the planets are initially positioned at a distance that ensures they are outside, yet near the Roche limit, the latter marking the relevant domain for which to begin using SPH, where the tidal forces should start to dominate over self-gravity. In order to make certain that our initial distance is always sufficiently large and outside the Roche limit, independent of the exact composition and density of every planet, we deliberately adopt an upper limit value in excess of the fiducial Roche values typical of rocky planets. Our selection is based on the analysis from \cite{VerasEtAl-2014} (see the discussion therein). The largest Roche value is given by their Equation 3 as $R_{roche,max}=2.73 R_\odot$. This value is derived from their Equation 2, by taking the upper range for the $C$ coefficient and the minimum permitted density of small asteroids. Were we to select both with average values instead, the Roche distance would have been at least halved. In short, given this choice we make sure that the SPH domain start of influence is always selected to be much larger than what is actually required, by a factor of at least 2. 

Since the \emph{actual} Roche limit is however less than 50\% smaller, we consider planet pericentre distances of around 1$R_\odot$ as having Roche-grazing orbits, yet well-placed inside the tidal sphere (i.e, at that distance they skim the Roche limit from within).

Throughout this section \emph{only}, we assign a small planet semi-major axis of merely 0.1 AU. The latter value is considered by \cite{VerasEtAl-2014} as the minimal value of $a$ for which the time spent inside the tidal disruption sphere is approximately independent of the choice of $a$. Additionally, this value is sufficiently low that the orbital period of the planet is only 14.9 days, which will allow us to track the formation of the disc for a considerable duration, of the order of 100-200 days, or $\sim10$ orbits. We note that such a small $a$ is physically highly unrealistic since most planetesimals are expected to originate from a semi-major axis of at least several AU (see discussion in Section \ref{SS:SPHOutline}), however this section is not meant to treat realistic scenarios, only maximize the simulation duration and characterize the resulting disc.

A typical simulation time of a few months (which translate into several orbital periods of the original planet), is achieved when the resolution is limited to 10K SPH particles, even when utilizing our relatively high-performance GPU-architecture. Taking a higher resolution must come at the cost of reducing the duration of the simulation (the fraction or number of orbits for which the disc formation is fully tracked) or lowering the semi-major axis/eccentricity (or both). Given our current choice of resolution, the typical runtime is coincidently comparable to the simulation time, most simulations running for up to 4 months. The simulations were performed on the 'TAMNUN' GPU cluster, at the Technion Institute in Israel. The GPU model used is NVIDIA Tesla K20. Each simulation ran on a single dedicated GPU.

\subsection{Dependence on pericentre distance}\label{SS:PericentreDependence}
In this Section we present the results of full-scale tidal disruption SPH simulations as a function of their depth, i.e., given the following pericentre values: 0.1$R_\odot$, 0.5$R_\odot$ and 1$R_\odot$. We consider the disruption of Mars (0.1$M_\oplus$) or Earth (1$M_\oplus$) sized planets. As discussed in the previous section, the planets are assigned a small semi-major axis of only 0.1 AU, for which their orbital periods is merely 14.91 days. In turn, this makes the tracking of disc formation computationally plausible when using full-scale SPH simulations.

\begin{figure*}
    \begin{center}	
        \includegraphics[scale=0.62]{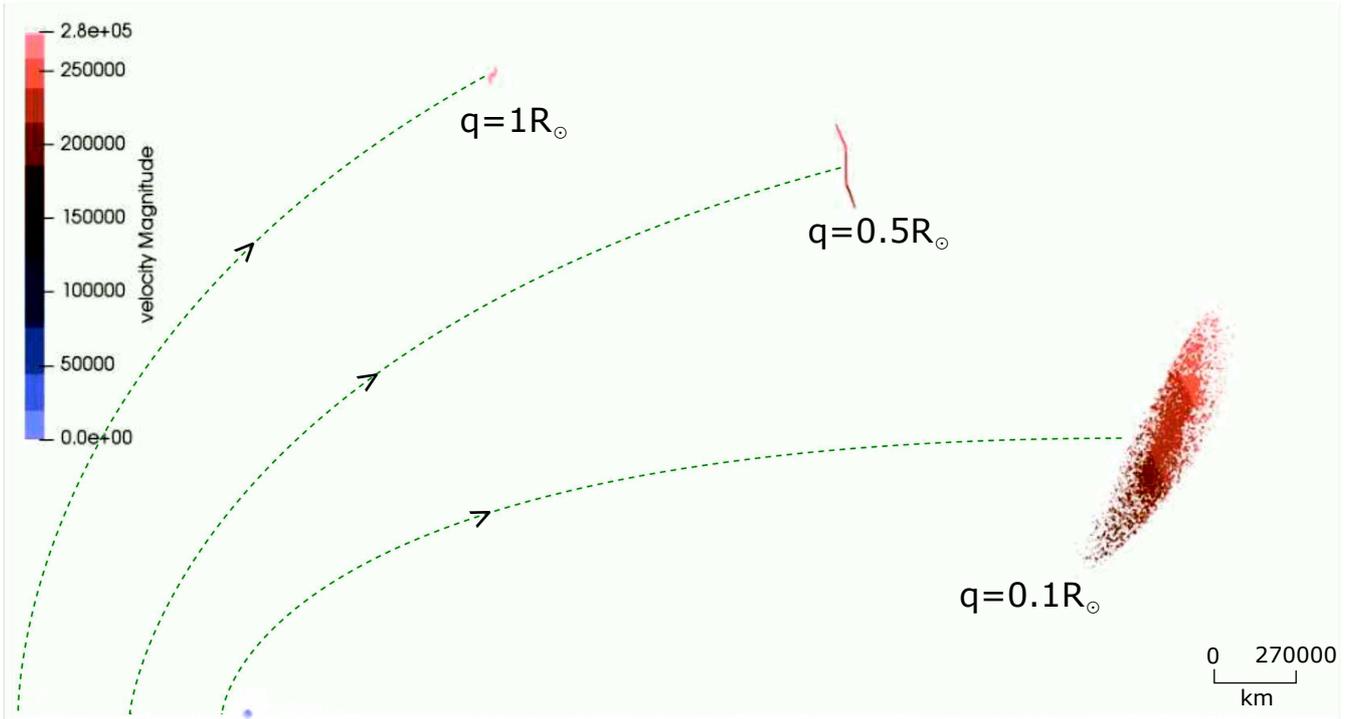}
 		\caption{A top-down snapshot of SPH particle velocities in units of m$\times$s$^{-1}$, after the initial tidal encounter ($\sim$13000 s). Three Earth-mass planets with $a=0.1$ AU and pericentre distances $q=$1, 0.5 and 0.1 $R_\odot$ (respectively from left-to-right) are shown moving clockwise (their trajectory lines are marked in dashed green), after they exit the tidal sphere of a 0.6$M_\odot$ WD (bottom-left, blue particle). The disruption causes the planet to distend, forming an outer and inner tidal streams. Deeper disruptions result in more complete tidal stripping. The velocity gradient corresponds to the transition from unbound to bound particles. The image is to scale.}
        \label{fig:VelRoche}
	\end{center}	
\end{figure*}

In Figure \ref{fig:VelRoche} we show the initial stage of tidal disruption. The image captures the debris after they exit the Roche limit, moving away from the WD. Three Earth-mass planets are considered with different $q$ values, from right-to-left: 0.1$R_\odot$, 0.5$R_\odot$ and 1$R_\odot$. Particle colours denote velocity (magnitude). Resolution is 10K particles. As can be seen, only the WD (bottom-left), represented by a single particle, is stationary. We recall from Section \ref{SS:SPHOutline} that the planets are initially positioned outside but near the Roche limit (by a factor of at least $\sim$2) on the opposite side of the WD. Their approximate trajectories are indicated by the green-dashed lines, moving clockwise. The time passed from their initial-to-final positions just outside this extended sphere, is around 0.15 days.

The disruption causes each planet to distend as tidal energy is transferred into the planet and matter crosses through the Lagrange points. Eventually two mass-shedding streams are evident. The inner stream particles are bound to the star, tracing out elliptical orbits, while particles in the outer stream are often unbound from both the star (and planet, if during a partial disruption it remains intact), heading away from the star system on hyperbolic orbits. Our simulations show that the streams generally follow a single axis, but the geometry obviously differs from case to case, depending on the distance of close tidal approach. In the $q=1R_\odot$ case, it is clearly evident, even during this early stage, that the mass shedding is partial and the streams emanating from the outer portions of the planet do not conform to a single axis geometry.

The velocities of the disrupted particles help us understand the initial formation of the disc. The general picture is as one might expect from Figure \ref{fig:roadmap}, the unbound particles along the tip of the outer stream having the highest velocities, whereas the particles further inward have increasingly lower velocities. Bound particles will slow until reaching their minimum, apocentre velocities. The slowest particles are positioned along the tip of the inner stream. They accordingly have the closest apocentre distances.

One of the major differences that emerge beyond this point is the degree to which the stream is gravitationally self-confined. Gravitational contraction is (by its definition) impossible while the debris still lie within the WD's Roche limit. However, as the particles continue to move away the gravitational interactions among them can, depending on their exact spatial distribution and velocities, cause them to clump up and form larger fragments. In other words, the stream may fragment under its own self-gravity.

Physical intuition regarding the fragmentation phase may be obtained based on the analysis of \cite{HahnRettig-1998}. We follow their calculations, in which they show that fragmentation may occur when the gravitational free-fall timescale $t_c$ becomes smaller than the stream spreading timescale $t_s$. The latter is calculated by determining the length $L(t)$ of the stream, over its rate of change $dL(t)/dt$. Using the notations from Section \ref{S:Analytical}, and replacing the breakup distance $d$ with the distance of close approach $q$, if the planet is on a highly eccentric orbit and $r_\mathrm{crit}\ll R$, then the most bound particle inside the inner stream has $\acute{a}=q^2/2R$. Manipulating the known relation $q=(1-e)a$, we obtain $\acute{e}=1-2R/q$. When $r_\mathrm{crit}\ll R$, the particle formerly at the planet's centre (i.e., with the orbital elements $a$ and $e$) approximately marks the other tip of the bound stream. Now $t_s$ can be calculated from the stream's length $L$ as a function of $t$ using standard solutions for elliptic Keplerian equations of motion, as follows. We calculate the mean anomaly $MA=2\pi t/T$, where $T=2\pi \sqrt{a^3/GM}$ is the orbital time (assuming $m\ll M$). The eccentric anomaly is obtained from solving Kepler's equation $MA(t)=EA(t)-e \sin{EA(t)}$ numerically. Then the distance as a function of time satisfies $l(t)=a(1-e\cos{EA(t)})$, the true anomaly $\theta$ satisfies $\theta(t)=2\arctan[\tan(EA(t)/2)*\sqrt{(1+e)/(1-e)}]$ and $l(t)_{x,y}$ can be extracted. From $\acute{a}$ and $\acute{e}$ we similarly extract $\acute{l(t)}_{x,y}$. Obtaining $L(t)$ and $dL(t)/dt$ is straight forward.

The former timescale $t_c$ is calculated based on the characteristic time it would take a cloud of mass to collapse under its own gravitational attraction. The free fall timescale equals $\alpha / \sqrt{G\rho}$, where $\alpha=\sqrt{3\pi/32} \cong 0.5427$ for a stationary cloud of particles. Since our tidal cloud of debris is not stationary, realistic values for $\alpha$ are larger, and must be calibrated from numerical simulations (e.g., in \cite{HahnRettig-1998} $\alpha$ is determined to be $\sim$1) in order to be used in the calculation. Since the debris spread largely along a single axis, the debris density $\rho$ equals the planet density, scaled by the factor $R/L$, such that $\rho \sim (3m/4\pi R^3) R/L$. According to \cite{HahnRettig-1998}, one has to equate $t_c$ and $t_s$, both which are time dependent, to find the moment $t$ when fragmentation becomes important. By the same analysis, we can also obtain an estimate regarding the number of fragments formed.

If $r_\mathrm{crit}\ll R$ is not satisfied, the calculation changes, however along the same general lines. One can use Equation \ref{eq:SemiMajorAxis} instead, to compute the interior bound orbit and the exterior bound orbit may be evaluated in the same way if $r_\mathrm{crit}>R$, otherwise $l(t)$ can be derived from the parabolic orbit equation. The main problem of the entire calculation is the calibration of $\alpha$, however as we shall soon see, it is not a trivial problem since the latter is actually a function of $q$. An additional caveat is that the method would fail to treat cases with large $q$, when the disruption is partial, since the temporal evolution of $L$ is different in this case and depends on the remaining intact planet.

\begin{figure}	
    \begin{center}		
 	    \subfigure[$q=1 R_{\odot}$, partial disruption.] {\label{fig:zoomin1}\includegraphics[scale=0.3]{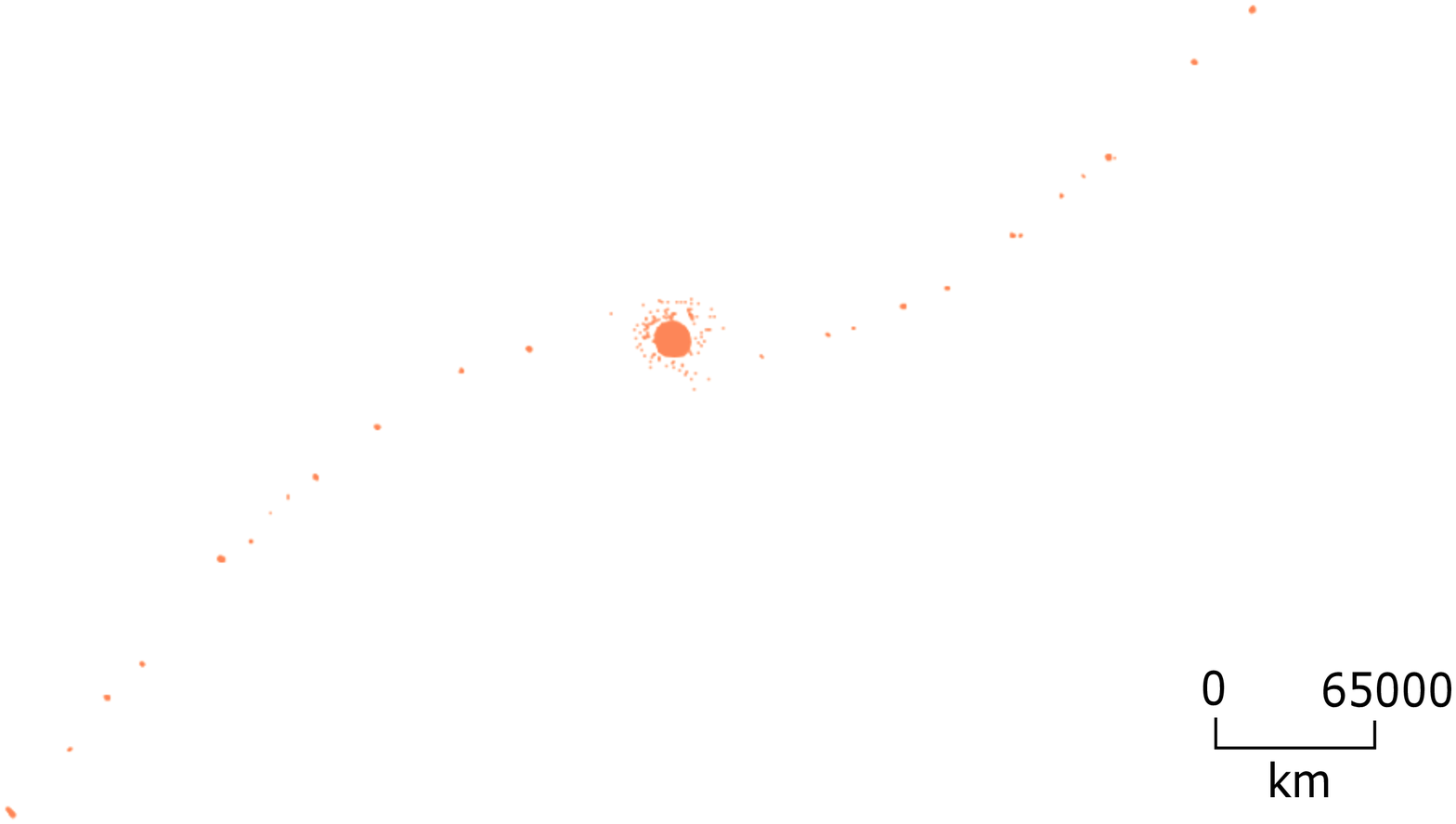}}
 		\subfigure[$q=0.5 R_{\odot}$, gravitationally self-confined full disruption. Zoom-in on dashed rectangle: Figure \ref{fig:fragMerger}.] {\label{fig:zoomin0_5}\includegraphics[scale=0.3]{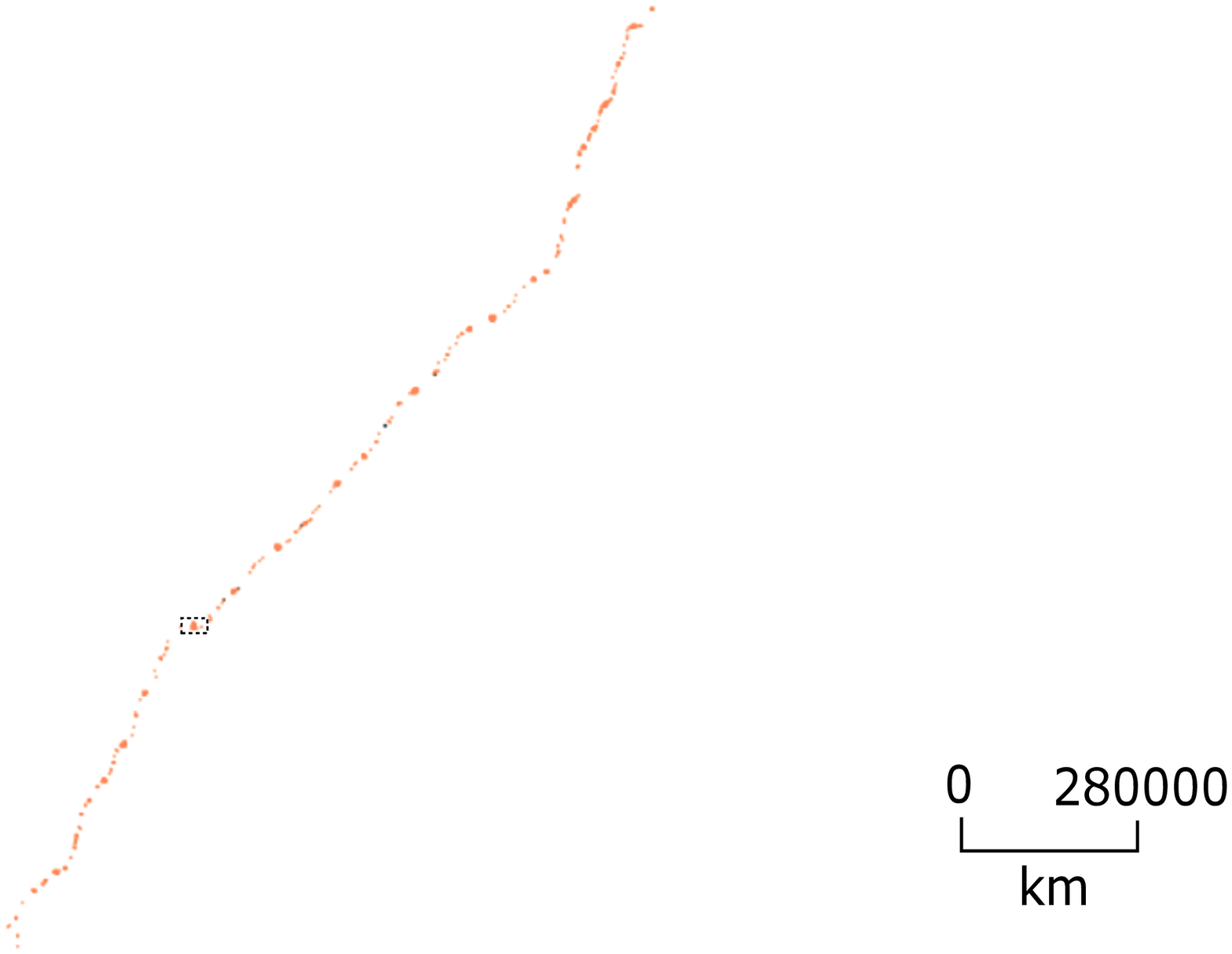}}
 		\subfigure[$q=0.1 R_{\odot}$, gravitationally unconfined full disruption.]
 		{\label{fig:zoomin0_1}\includegraphics[scale=0.3]{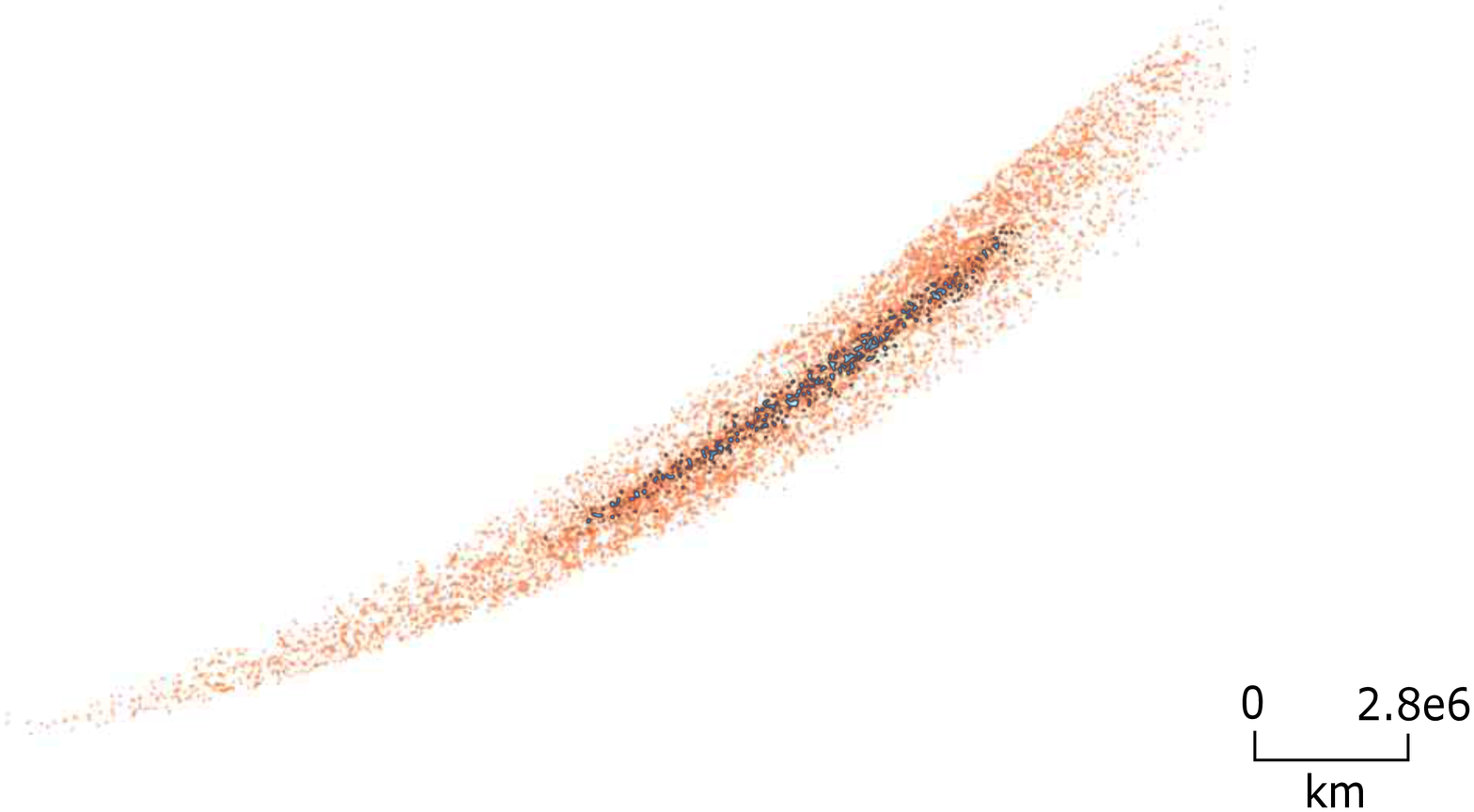}}		
 	\end{center}
 	\caption{The zoomed-in debris field (top-down view) of a tidally disrupted Earth-sized planet around a 0.6$M_\odot$ WD after 0.58 days, given $a=0.1$ AU and a $q$ of (a) 1$R_{\odot}$; (b) 0.5$R_{\odot}$ and (c) 0.1$R_{\odot}$ (panel dimensions not identical). Colour denotes composition: orange - rock ; black - iron.}
    \label{fig:zoomin}
\end{figure}

Thus, considering intermediate or deep disruptions, our initial approach to this problem was to plainly rely on our numerical simulations to find out the exact time when fragmentation occurs, by visually inspecting our data for several scenarios. Based on our comprehensive analysis of full SPH simulations, we found that for $q=0.1R_\odot$, the SPH particles do not coalesce into fragments at all. For $q=0.5R_\odot$, the typical timescale for fragmentation is of the order of a few $10^-1$ days. In Figure \ref{fig:zoomin} we zoom-in on the debris after this fragmentation phase has concluded. Taking the aforementioned timescale estimate, our time index is now 0.58 days. Pixel colours denote composition: orange - rock; black - iron. Resolution is 10K particles.

Our choice of $q$ highlights three distinct cases. Unlike in Figure \ref{fig:VelRoche}, the particles no longer form a continuous stream. It is visually evident that the amount of mass stripped from the planet increases with decreasing periastron separation. Panel \ref{fig:zoomin1} shows a clear case of a partial disruption, in which a relatively intact planet, is accompanied by a small stream of particles from its outer portions. Panel \ref{fig:zoomin0_5} breaks up the planet entirely into a long and narrow stream, but the debris field is gravitationally self-confined, and the stream coagulates to form a finite number of large fragments, accompanied by some smaller particles. Panel \ref{fig:zoomin0_1} showcases the deepest, most violent type of disruption, wherein the destruction of the planet is almost complete, in addition to the debris being so dispersed that they are unable to fragment under the pull of their own self-gravity.

\begin{figure}
	\begin{center}	
 		\includegraphics[scale=0.558]{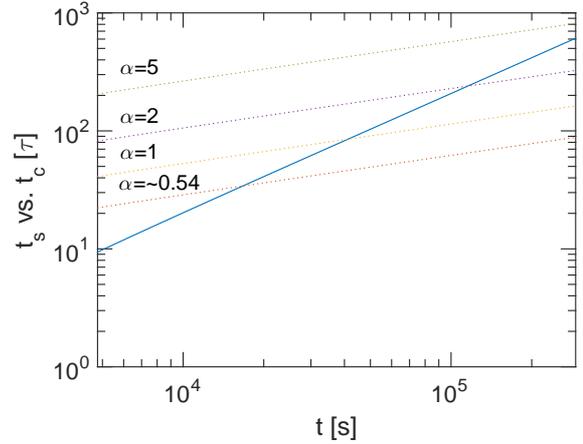}
 		\caption{The tidal stream spreading timescales $t_s$ (solid line) and free-fall timescale $t_c$ (dotted lines, with various $\alpha$ parameters) in units of the encounter timescale $\tau$, as a function of $t$ (both in logarithmic scale). When $t_c$ is less than $t_s$ gravitational contraction and consequently fragmentation, is possible.}
 		\label{fig:FragTimescale}
	\end{center}	
\end{figure}

Let us now examine the fragmentation timescale using the \cite{HahnRettig-1998} approach, comparing our results to theory. After performing the previously mentioned calculations, we present in Figure \ref{fig:FragTimescale} the timescales $t_s$ (solid line) and $t_c$ (dotted lines) in units of the encounter timescale $\tau=\sqrt{q^3/GM}$, and in logarithmic scale. To compare with Figure \ref{fig:zoomin}, here we also perform the calculation for an Earth-mass planet with $a=3$ AU. The x-axis shows the time $t$. The starting time ($\sim 5\times 10^3$ s) corresponds to a position that lies beyond the maximum Roche limit (see e.g. \cite{VerasEtAl-2014}), and the end time equals $t=T/2$, where $T/2$ is half the orbital period of the innermost bound particle in the stream. The latter choice is significant since it is the moment in which particles begin to deviate from the single axis geometry of the stream. Within this critical time interval, the fragmentation timescale is obtained when the free-fall timescale $t_c$ intersects with the spreading timescale $t_s$, while we investigate several values of $\alpha$, spanning one order of magnitude.

It is shown that for $\alpha=1$, gravitational contraction is possible after about 40000 s, or 0.46 days. It is a good match to our numerical simulations and visual inspection of the data, for $q=0.5R_{\odot}$ (and also a good agreement with \citep{HahnRettig-1998}). The fact that no fragmentation ever occurs for $q=0.1R_{\odot}$, suggests that $\alpha$ is inversely correlated with $q$. In other words, the theoretical interpretation of our results might suggest that a large $\alpha$ inhibits the onset of fragmentation, and merits future investigation in these directions (see Section \ref{SS:Fragmentation}).

In order to quantify the effects of fragmentation, we analyse the data for the number of fragments (we find physical clumps of spatially connected SPH particles whose distances are less than the simulation smoothing length, using a friends-of-friends algorithm), after the particle fragmentation phase has concluded. Our analysis indicates that for Earth-sized planets, approximately 91.7\% of the particles are single SPH particles and are not associated with any fragment, when $q=0.1R_\odot$. In other words, when the disruption is so deep these planets mostly break down into their smallest constituent particles (here limited by resolution), and are unable to fragment significantly. The other 8.3\% of the particles are distributed among small fragments, many of which include only a few particles. In comparison, for $q=0.5R_\odot$ and $q=1R\odot$, the fractions of single SPH particles are much smaller, only 1.88\% and 0.54\% respectively.

A Mars-sized planet (not shown in Figure \ref{fig:zoomin}) breaks down into fewer single SPH particles compared to an Earth-sized planet. Our analysis shows that the respective fractions of single SPH particles in that case are only 76.5\%, 0.74\% and 0.05\%.

\begin{figure*}	
	\begin{center}		
		\subfigure[$t=0.58$ days.] {\label{fig:fragMerger1}\includegraphics[scale=0.475]{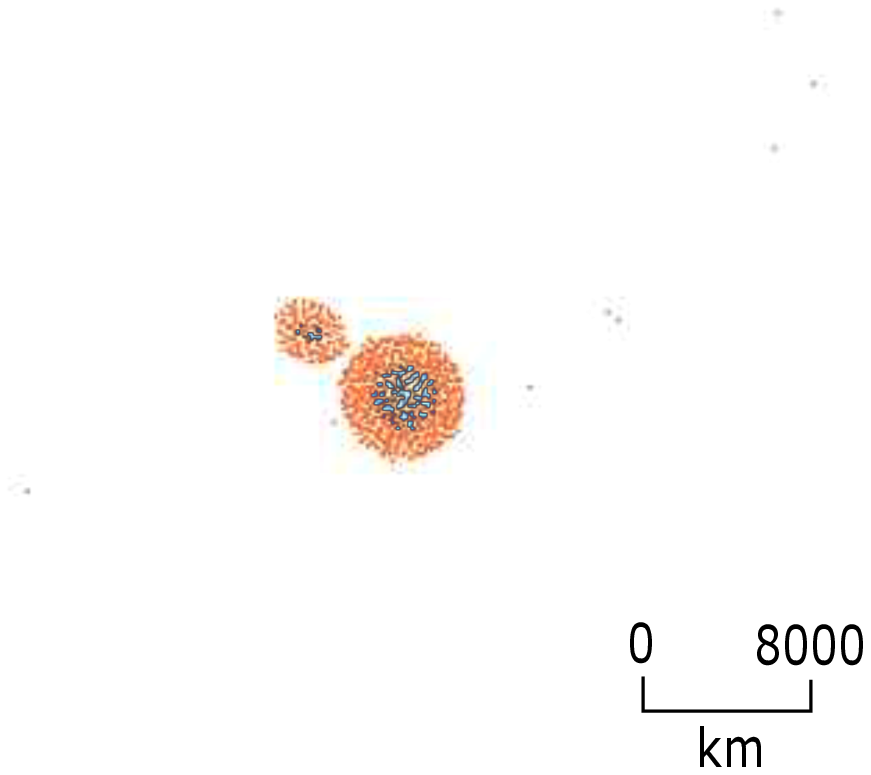}}
		\subfigure[$t=0.64$ days.] {\label{fig:fragMerger2}\includegraphics[scale=0.475]{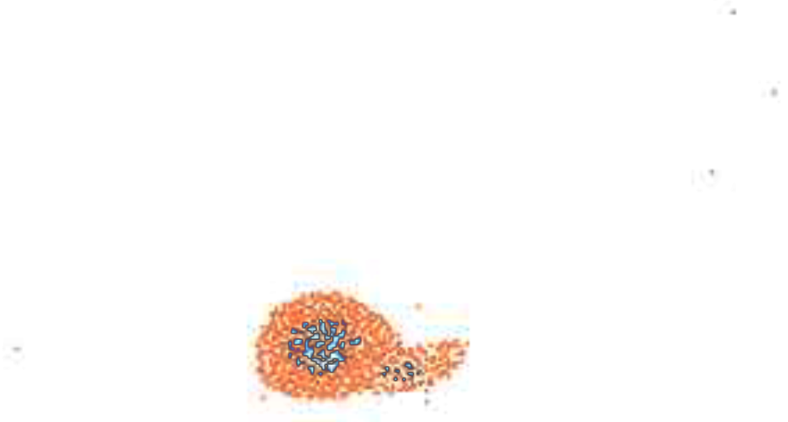}}
		\subfigure[$t=0.7$ days.]
		{\label{fig:fragMerger3}\includegraphics[scale=0.475]{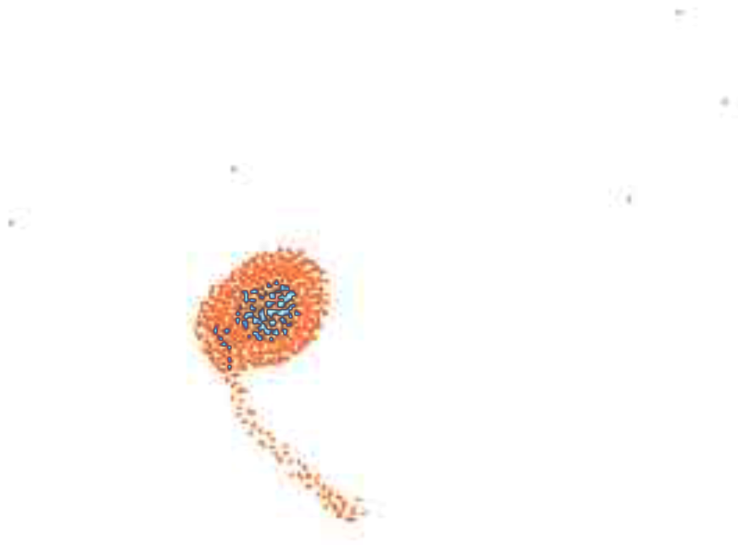}}
		\subfigure[$t=0.76$ days.] {\label{fig:fragMerger4}\includegraphics[scale=0.475]{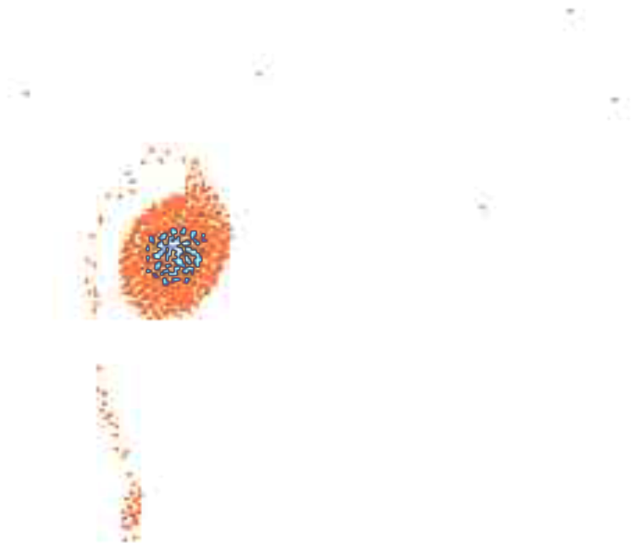}}
		\subfigure[$t=0.82$ days.] {\label{fig:fragMerger5}\includegraphics[scale=0.475]{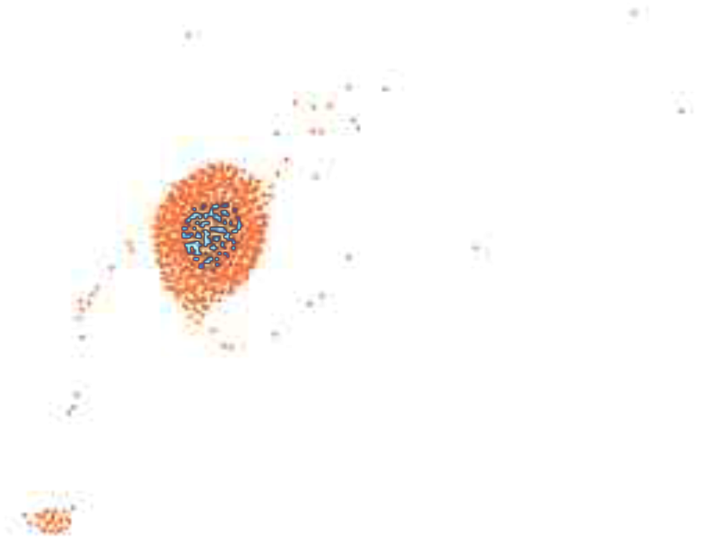}}
		\subfigure[$t=1.16$ days.]
 		{\label{fig:fragMerger6}\includegraphics[scale=0.475]{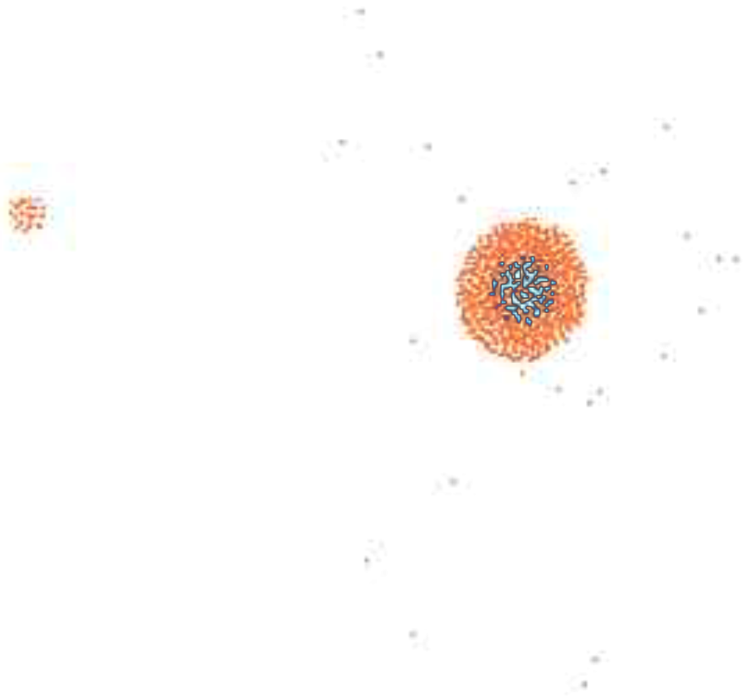}}		
	\end{center}
	\caption{Zoom-in on fragment dynamics inside the gravitationally self-confined stream from Figure \ref{fig:zoomin0_5} (see dashed rectangle therein). Scale in Panel \ref{fig:fragMerger1}. This top-down view shows two fragments collide and merge, producing a secondary field of smaller debris. At 0.58 days, the start time is identical to Figure \ref{fig:zoomin}, and ending at 1.16 days. Colour denotes composition: orange - rock ; black - iron.}
	\label{fig:fragMerger}
\end{figure*}

The composition distribution is easily observed in Panel \ref{fig:zoomin0_1}. Here the iron particles from the planet's core experience the same tidal shearing as the rocky particles, and display a similar spatial pattern. Panels \ref{fig:zoomin0_5} and \ref{fig:zoomin1}, are however opaque and insufficiently zoomed-in for visual identification. A much closer inspection would nonetheless reveal that in Panel \ref{fig:zoomin0_5} the iron particles are distributed inside the cores of (many of) the larger fragments, as in, e.g., Figure \ref{fig:fragMerger}. In Panel \ref{fig:zoomin1} a closer inspection shows that the iron particles are found \emph{only} inside the core of the original, intact planet, whereas all other small fragments are purely rocky.

It is interesting to note in this context, that the rocky interstellar asteroid Oumuamua is sometimes said to be an unbound fragment originating from a tidally disrupted planet \citep{Cuk-2018,RaymondEtAl-2018a,RaymondEtAl-2018b}. \cite{Rafikov-2018} went one step further in postulating that it could originate from a disruption around a WD, specifically, since polluted WDs often showcase a characteristic abundance of refractory materials. He introduced a complex model for producing the fragment size distribution required for the small size of Oumuamua, by collisional grinding of fragments which arises during the planetary passage through the Roche sphere. In this study, our results are reinforcing the notion that objects like Oumuamua, being so small, must also originate in one of two formation pathways. The first option is that they form specifically in gravitationally unconfined streams, otherwise the outgoing stream would coagulate into larger fragments as it exits the Roche limit, regardless of how small the pieces are when they initially form. This generally means that the disruption has to be very deep. The other option, when we have not-so-deep, yet full disruptions, and the stream is gravitationally self-confined, is to form objects like Oumuamua as smaller, second-generation particles. Close inspection of our data reveals that the way in which to do that is through collisions among merging fragments.

As a demonstration, let us continue following the gravitationally self-confined stream from Figure \ref{fig:zoomin0_5}. After $t=0.45-0.55$ days, all particles have conjoined to form fragments. However, we then see some adjacent fragments, that are gravitationally interacting with one another and eventually merging. Figure \ref{fig:fragMerger} shows an example of such a collision. We note that it is by no means unique, either in this particular simulation, or in our entire suite of simulations. We see many such mergers during this (henceforward-termed) collisional phase, and the transfer of angular momentum in such collisions often results in fast rotation and shedding of mass, producing a cloud of smaller debris in orbit around the central, rotationally-oblate fragment. Sizeable satellites also occasionally form, as in Figure \ref{fig:fragMerger6}, however just as often the second-generation debris field is composed of merely small fragments and a lot single SPH particles. We note that the minimum particle size here is resolution-limited, however there is no physical reason to assume that the secondary debris cloud is not composed of yet smaller particles than our resolution permits, potentially following a power law size distribution, including tiny satellites, boulders and dust.

Following the collisional phase, the stream settles into an henceforward more stable and collisionless phase, in which the fragments remain mostly unaltered, at least until the next time they enter a strong gravitational potential (like the star, if they are still bound to it). Our inspection of the data seems to suggest this phase ends at roughly 1.16 days, hence it has a duration similar to that of the fragmentation phase.

\subsection{Disc formation}\label{SS:DiscFormation}
\begin{figure}	
	\begin{center}		
		\subfigure[1st pass, $t=1.16$ days] {\label{fig:1pass}\includegraphics[scale=0.43]{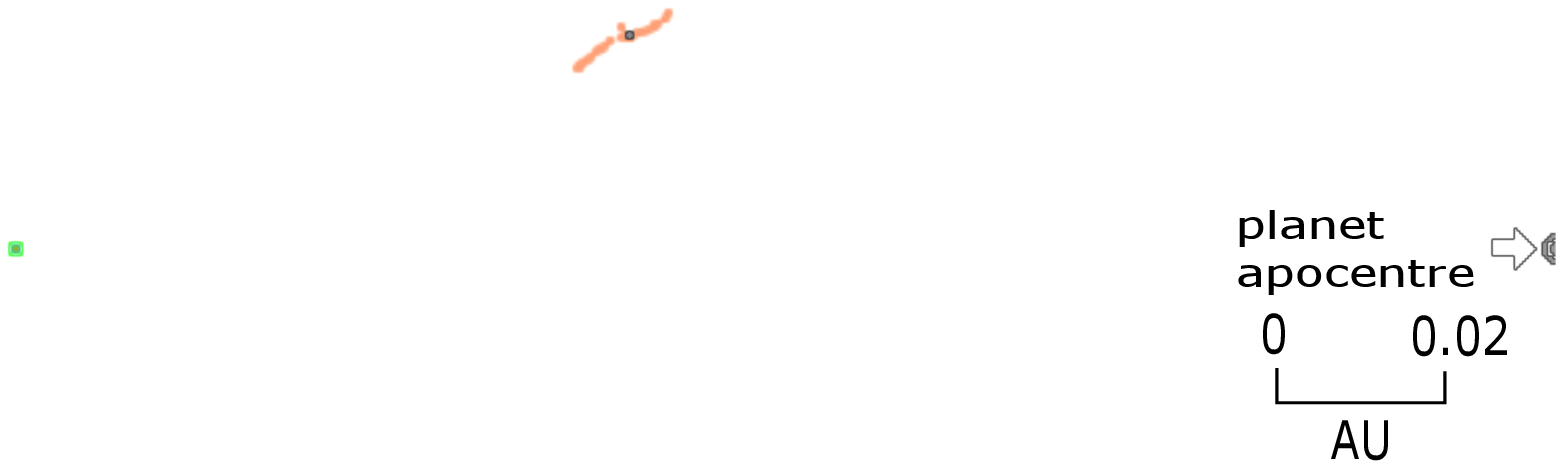}}
		\subfigure[2nd pass, $t=16$ days]
		{\label{fig:2pass}\includegraphics[scale=0.43]{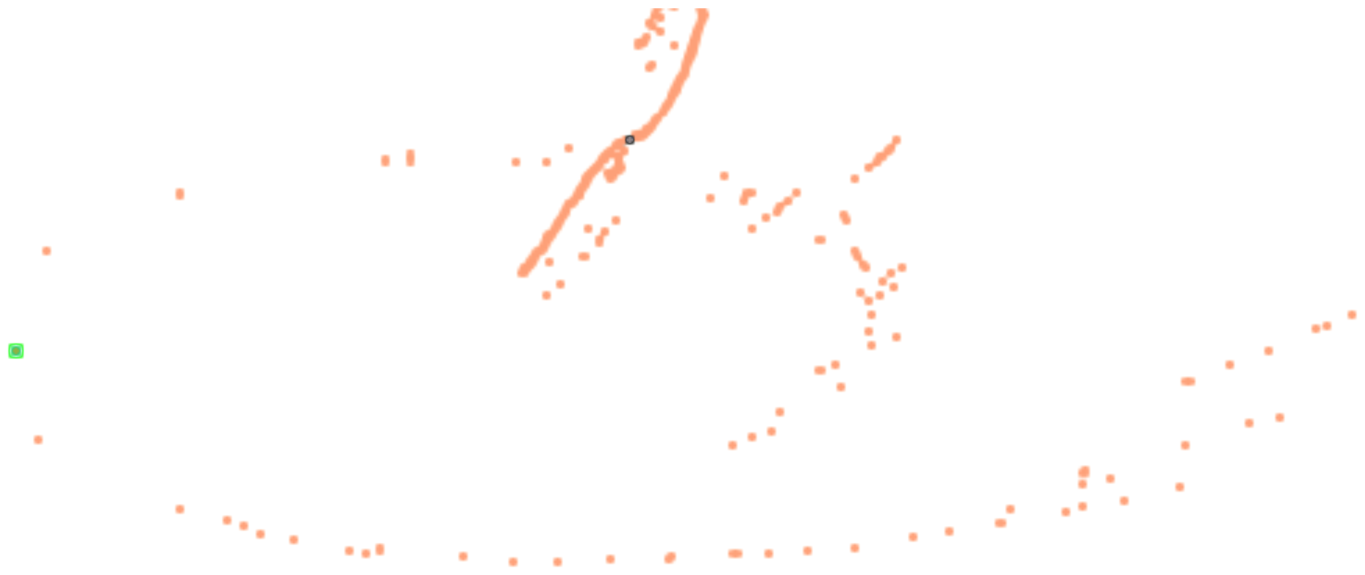}}
		\subfigure[3rd pass, $t=31$ days] {\label{fig:3pass}\includegraphics[scale=0.43]{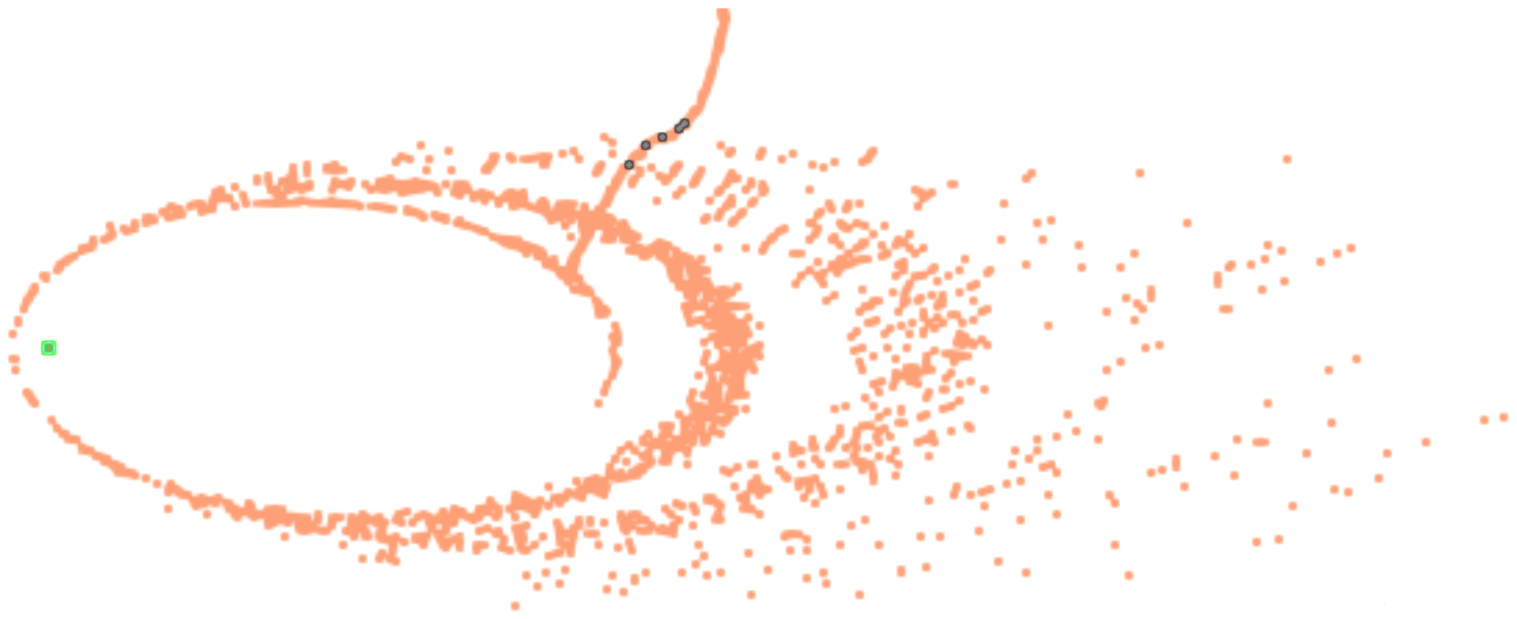}}
		\subfigure[4th pass, $t=46$ days] {\label{fig:4pass}\includegraphics[scale=0.43]{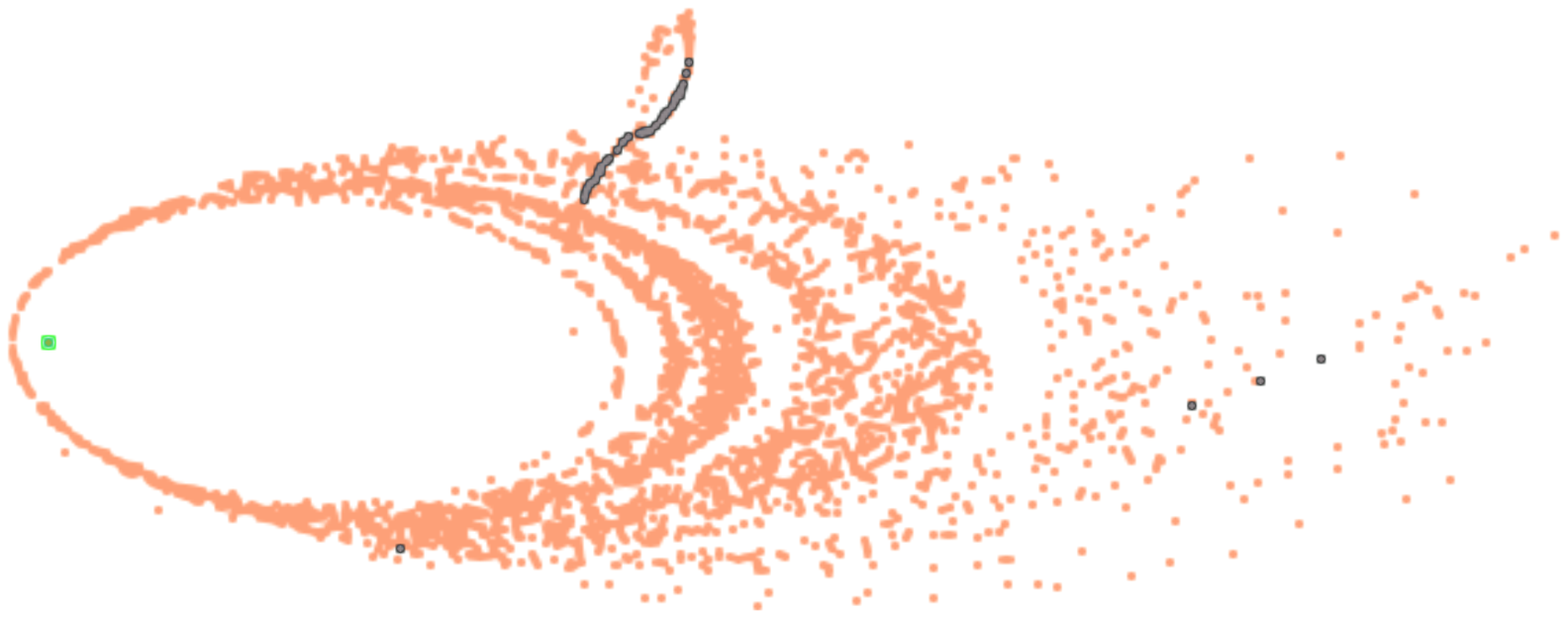}}
		\subfigure[5th pass, $t=61$ days]
 		{\label{fig:5pass}\includegraphics[scale=0.43]{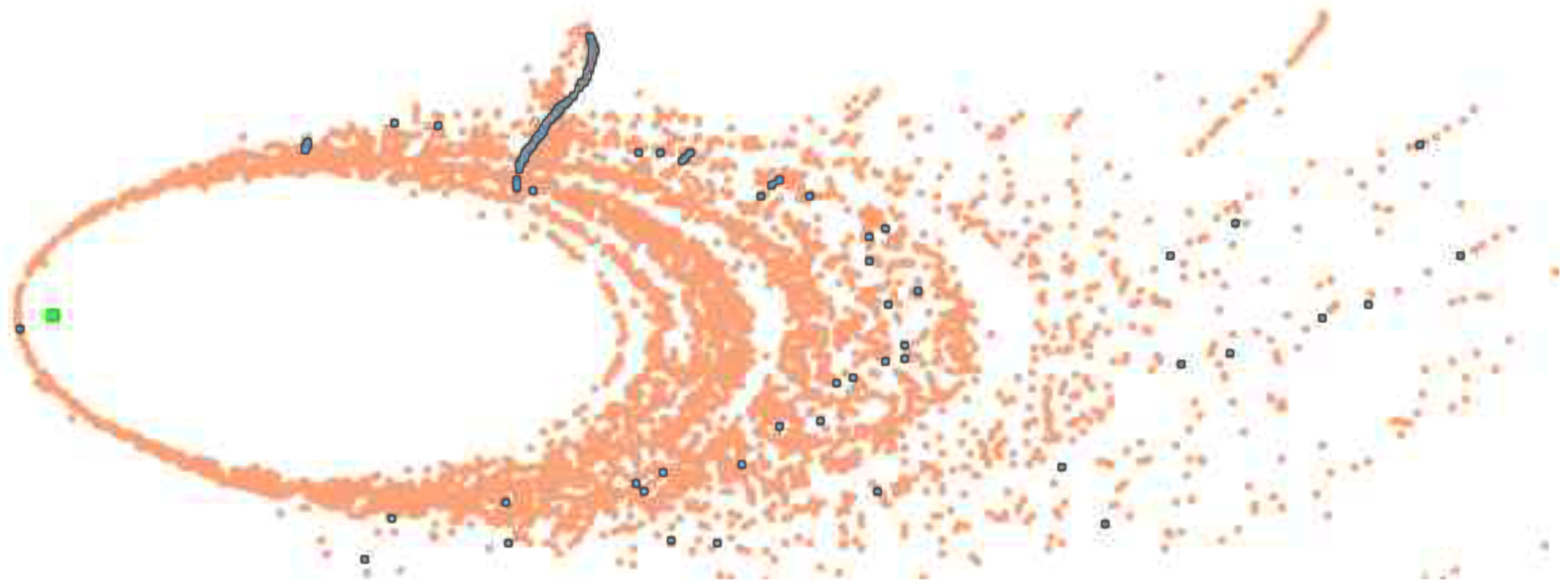}}
		\subfigure[10th pass, $t=119$ days]
 		{\label{fig:10pass}\includegraphics[scale=0.43]{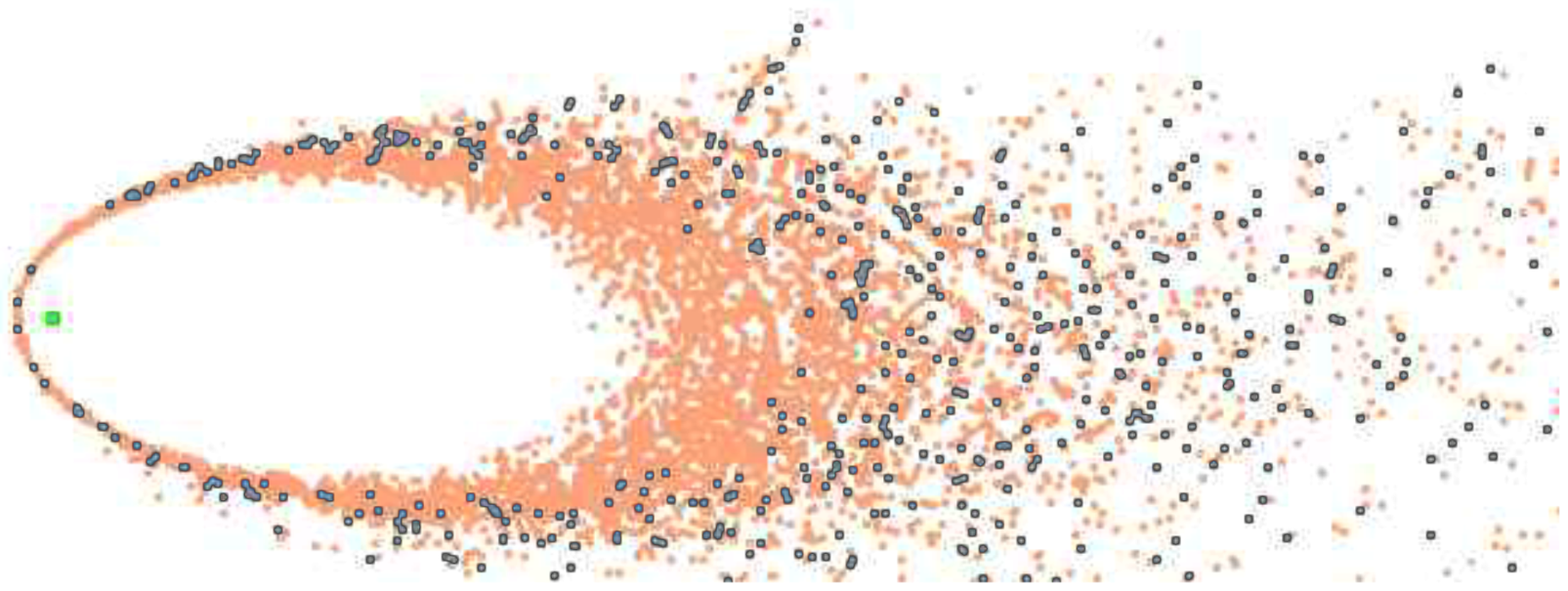}}
	\end{center}
	\caption{Top-down view of disc formation through a sequence of partial tidal disruptions of an Earth-sized planet with $a=0.1$ AU and $q=1 R_{\odot}$, around a 0.6$M_\odot$ WD, plotted at different multiples of the original orbit. Colour denotes composition: orange - rock ; black - iron; green - WD. The planet apocentre and scale is indicated in Panel \ref{fig:1pass}.}
	\label{fig:EarthDisruption}
\end{figure}

The formation of the disc continues as each returning fragment completes a full orbit (given its new semi-major axis) and re-enters the tidal sphere. Although each fragment has a different size, composition and semi-major axis compared to the planet from which it stem, its pericentre distance does not change. It may therefore tidally disrupt again during close approach, further breaking into smaller and smaller fragments, and so forth. If and when a fragment disrupts, we get a dispersion in the semi-major axes of resulting sub-fragments. This process is visualized in Figure \ref{fig:EarthDisruption}, where we show the disc formation progress of a tidally disrupted Earth-sized planet with a pericentre of $q=1 R_{\odot}$ (and $a=0.1$ AU, as before). Due to its large pericentre distance, the planet only grazes the Roche limit and thus essentially undergoes only partial disruptions during each pass, typically shedding a smaller fraction of mass compared to deeper disruptions. Such a disc is the slowest to evolve, and we will follow it here in discrete time intervals equal to the original planet orbital period of 14.9 days. Colour denotes composition: orange - rock ; black - iron; green - WD. Resolution is 10K particles.

In Panel \ref{fig:1pass} the first pass through the tidal sphere is shown, at t=$10^5$ (here we wait twice longer than the fragmentation time, for the tidal streams to distend further, in order to get a better visual effect). The planet remains almost fully intact, however an outer and inner tidal stream develop, composed of several small rocky fragments, as in Figure \ref{fig:zoomin1}. The planet continues to move on its original trajectory, its apocentre located to the right edge of the frame, as indicated by the arrow and markings. The fragments at the tip of the inner tidal stream will be the first to reach their new, smaller apocentres.

Continuing to Panel \ref{fig:2pass}, we follow the progression of the disc as it evolves. The aforementioned rocky fragments located at the tip of the inner stream that were formed during the first tidal approach in Panel \ref{fig:1pass}, had an apocentre distance only half of that of the original planet. By $t=16.07$ days they have already re-entered the tidal sphere, disrupting on their own, each creating once again a less massive tidal stream (with its own, new dispersion in particle semi-major axes). This sequence of events begins to form interlaced elliptic eccentric annuli, each resulting in a different angular size depending on the physical and orbital properties of the original fragment from which they were produced. The smallest fragments form eccentric rings instead. 

Focusing again on the planet, it now undergoes its second disruption, which highly resembles the first. However, this disruption sheds much more mass from the planet's rocky exterior, an outcome which we explain by the planet's rapid spin, as follows. Our analysis shows that the planet has obtained a 3.3 h rotation period during the first disruption. Tidal spin-up is a well known phenomena, which has been previously seen in simulations involving soft tidal encounters for a wide range of applications, including stars \citep{AlexanderKumar-2001,AlexanderKumar-2002} and asteroids \citep{RichardsonEtAl-1998,WalshRichardson-2006,MakarovVeras-2019}. To the best of our knowledge, however, our paper is the first to report and (in Paper II) statistically analyse the tidal spin-up of terrestrial planet fragments.

We will show that self-rotation of the planet yields a larger stellar Roche limit, effectively making the tidal disruption of the planet deeper, given the same approach distance as before. In order to illustrate this point, consider the simple, classical derivation of the Roche limit. Assuming no rotation, one equates the force of self gravity $F_{sg}$ exerted by the planet with the tidal force $F_T$ exerted by the star. Using the same notation as in Section \ref{S:Analytical}, this gives the standard $R_{roche}=R*(2M/m)^{1/3}$, which we can also flip to obtain $R=R_{roche}(m/2M)^{1/3}$. Now Equation \ref{eq:TidalForce} can be re-written as a function of $R_{roche}$:

\begin{equation}
F_T=\frac{G(2M)^{2/3}m^{1/3}R_{roche}}{d^3}
\label{eq:TidalForce2}
\end{equation}

By adding the planet's rotation, however, we now have $F_{sg}=F_T \pm F_{rot}$. The negative sign before $F_{rot}$ is applicable to retrograde rotation. Since here the planet's rotation is excited during the initial tidal approach in the same sense as its orbit, we take only the positive sign. We thus have:

\begin{equation}
\frac{Gm}{R^2} = \frac{2GMR}{R_{roche}^3} + \omega^2 R
\label{eq:RocheWithRot1}
\end{equation}

\noindent where $\omega$ is the planet's rotation rate. It is convenient to express $\omega$ in terms of the planet's breakup rotation $\omega_{br}=\sqrt{Gm/R^3}$ (which is obtained by equating the forces of self-gravity and self-rotation), such that $\omega=\lambda \omega_{br}$, $\lambda$ being the breakup velocity fraction. Rearranging Equation \ref{eq:RocheWithRot1} and solving for $R_{roche}$ we get:

\begin{equation} 
R_{roche} =  R \left( \frac{2M}{m}   \right)^{1/3}   \left( 1-\lambda^2    \right)^{-1/3}
\label{eq:RocheWithRot2}
\end{equation}

We note that when $\lambda\ll 1$, Equation \ref{eq:RocheWithRot2} recovers the standard Roche limit, while for $\lambda=1$, $R_{roche}$ goes to infinity, as one expects.

Since $R_{roche}$ increases due to self-rotation, it follows that the same close approach distance $d$ is now comparatively deeper than in the non-rotating case. Plugging Equation \ref{eq:RocheWithRot2} into Equation \ref{eq:TidalForce2}, the tidal force effectively increases by a factor of $(1-\lambda^2)^{-1/3}$. Given the Earth's breakup rotation period (approximately 1.4 h), and the 3.3 h rotation period of the returning planet from Panel \ref{fig:2pass}, we have $\lambda \cong 0.424$ and $\lambda^2 \cong 0.18$. Thus the tidal force effectively increases by about 7\%. Note that in this simple analysis, we ignored the planet's significant ellipsoidal shape due to fast rotation, which, as we recall from Equation \ref{eq:TidalForce}, increases the tidal force even more, and at the same time reduces self-gravity on its surface. Hence, the balance between these two forces is offset even more. We nevertheless note that the tidal force magnitude still remains dominated by the distance of close approach $d$. Self-rotation induces a much smaller effect, yet it facilitates more mass shedding, especially when $\lambda$ is large. Even when $\lambda$ is small, self-rotation can modify the energy dispersion in the stream, as we discuss in Appendix \ref{A:RotationDependence}. 

Moving to Panels \ref{fig:3pass}-\ref{fig:10pass}, the same qualitative behaviour continues. As mass is subsequently being shed from the planet with each tidal approach, $\omega_{br}$ increases with decreasing size, but so does $\omega$ (which is spun up during the disruptions). In Panel \ref{fig:3pass} a few small iron fragments break off from the planet, each consisting of several particles. In Panel \ref{fig:4pass} the streams include numerous fragments with the size of small asteroids, each with a tiny iron core and outer rocky layer, however the planet is still intact. The main change occurs in Panel \ref{fig:5pass}, where the fast-spinning planet breaks into a chain of multiple large fragments, as in Figure \ref{fig:zoomin0_5}, albeit these fragments are now largely composed of iron, as most of the rocky material has already been removed during the previous tidal approaches. Finally in Panel \ref{fig:10pass} we obtain the eventual properties of the fully formed disc. We expect very few changes to occur beyond this point. There remain only a few bound fragments that were flung to distances beyond the original semi-major axis, which will return for subsequent passes and disruptions around the star. By numbers they represent a negligible fraction and may be ignored.

Interestingly, the inner part of the disc is solely rocky, which seems a recurring feature in many of our simulations. This could be intuitively understood from the fact that in Panel \ref{fig:2pass}, where we have the first major disruption, the planet is larger than in \ref{fig:5pass}, in which the iron finally enters the tidal streams during the last cataclysmic disruption. In both cases the planet has the same $r_\mathrm{crit}$ (same $a$ and $d$), which means that the former disruption is more dispersive than the latter (smaller $\acute{a}$ due to larger $R$), explaining why the inner part is solely rocky.

\section{The hybrid approach}\label{S:Hybrid}
As previously shown, the complete disc formation always requires a very long tracking time and typically multiple and repeated disruptions. In Section \ref{S:FullSPH} our approach in handling this computational problem was similar to all other previous studies. We deliberately chose unrealistically small orbits and a low resolution for our simulations. Real tidal disruption scenarios are however often characterized by extremely large and eccentric orbits, making them computationally inaccessible to any numerical method proposed until now. The hybrid approach described in this section does not suffer from the same limitation. It enables to track the tidal disruption and disc formation of any planet, regardless of how far away its orbit is, with exactly the same efficiency. This level of performance does come with a price, since it entails certain assumptions. However, in the following section we will show that these are negligible compared to the potential gain.

\subsection{Principles and assumptions}\label{SS:HybridDescription}
As we have shown in Section \ref{SS:PericentreDependence}, extremely deep tidal disruptions give birth to violent and gravitationally unconfined streams. They break up the planet into its constituent particles and prevent the further coagulation of larger fragments. The outcomes of such disruptions can be fairly-well characterized analytically, and in some cases can also be tracked fully with SPH within reasonable timescales. We therefore note that the hybrid method was never intended to treat these cases, although it can certainly do so, and even with superior performance.

The hybrid approach is intended for an efficient treatment of partial disruptions, or full disruptions which are not as deep, and result in gravitationally self-confined tidal streams. The approach makes use of the fact that the primary processes taking place during these tidal disruptions, are restricted to a relatively small spatial domain. First, we recall that the differential gravitational force that breaks the object apart, is relevant only to the Roche limit of the star. The second important phase is fragmentation. It is during this phase, that small particles may collapse by the stream's own self-gravity, to form larger fragments. The relevant spatial domain here, as discussed in Section \ref{SS:PericentreDependence}, is also rather small, exceeding the tidal sphere only by an order of magnitude or so. Overall, the breakup and fragmentation phases constitute only a tiny fraction of the total spatial domain (of the original orbit), and are confined to the immediate environment of the star.

Our approach is therefore to restrict the SPH computations only to this relatively small domain, and to omit unnecessary calculations outside of it. Following the fragmentation phase, we identify the emerging fragments (whose constituent particles form spatially connected clumps of material), and for the reminder of their orbits, their trajectories are calculated and tracked analytically, assuming Keplerian orbits. Our hybrid program simply places each fragment once again near the star's Roche limit, based on its return orbital elements. This instantaneous 'trasport' comes with the price of ignoring the possible interactions or collisions of this fragment with other fragments or pre-existing material orbiting the star (see discussion in Appendix \ref{A:Collisionless}), in addition to other processes like radiation effects from the star (Appendix \ref{A:Radiation}). However, it saves a lot of computational time, since only a small fraction of the orbit (e.g. $\sim 10^{-2}$ even in contrived, low eccentricity test cases such as in Figure \ref{fig:SPHVsHybrid}) is simulated in full. The returning fragment immediately undergoes an additional tidal disruption, potentially splitting into a new set of fragments, with its own unique dispersion in orbital parameters, and so on. The discussion in Appendices \ref{A:Collisionless} and \ref{A:Radiation} shows that the partitioning of the tidal disruption problem is judicious in this case.

The hybrid code's main task is to identify the fragments, accurately calculate their orbits and especially handle the synchronization and timing of the subsequent disruptions. Apart from this, its other procedural task is handling the SPH job dissemination. The hybrid code terminates when reaching one of two outcomes: either all fragments have ceased disrupting given their exact size, composition and orbit; or fragment disruption is inhibited when reaching its minimum size - that of a single SPH particle.

\subsection{Code implementation}\label{SS:HybridImplementation}
Our code performs the following sequence of steps: 

\noindent\textbf{(a)} Based on the input orbital parameters of the disrupting planet, and the mass of the star, a position outside twice the Roche limit and the planet orbital period are both calculated.

\noindent\textbf{(b)} Based on the initial radial distance and orbital period, the initial true anomaly, position and velocity of the planet are calculated.
 
\noindent\textbf{(c)} The required SPH-duration to achieving fragmentation is calculated (the collisional phase is neglected to get a factor 2 improvement in computation speed). 

\noindent\textbf{(d)} The needed input files for the \emph{miluphCUDA} SPH code are generated, including a pre-processed relaxed planet for initializing the SPH disruption simulation.

\noindent\textbf{(e)} Excute SPH job on GPU.

\noindent\textbf{(f)} Analyse result, generate new SPH input files and repeat step (e) until no further disruptions occur or 100\% of the fragments have been reduced to single SPH particles.

\noindent\textbf{(h)} Finalize output/visualization files. \\
	 		
Steps (a)-(h) are scripted with BASH. Jobs are disseminated via a linux portable batch system (PBS). The main body of code (approximately 1500 lines) is carried out in step (f) via a separate C program, performing these steps:

\noindent\textbf{(f1)} Find physical fragments (clumps) of spatially connected SPH particles using a friends-of-friends algorithm.

\noindent\textbf{(f2)} Compute fragment properties, in addition their centre of mass (COM) positions and velocities.

\noindent\textbf{(f3)} Compute fragment orbital elements and subsequent disruption times (returning fragments are assumed to be transported to exactly the same distance from the star, regardless to any change in their composition. That is why we calculate a position outside twice the estimated rocky Roche limit).

\noindent\textbf{(f4)} Sort fragments (and their inherent particles, i.e. their relative positions and velocities with respect to the fragment COM) by their subsequent disruption times.

\noindent\textbf{(f5)} Generate the input files to start the subsequent SPH job (i.e., the disruption of the next fragment, after performing the appropriate synchronization procedures).

\noindent\textbf{(f6)} Generate visualization output files, between the current time and the next tidal disruption time (or simulation time limit, whichever smaller).

\subsection{Performance and validation of the hybrid model}\label{SS:Hybrid_Performance_Validation}
Our goal in this section is to test the hybrid SPH-Analytical model against similar models that were carried out in full, using only SPH. If successful, the hybrid simulation should produce an identical disc of debris, but it will achieve this goal in significantly less time.

\begin{figure*}	
	\begin{center}		
		\subfigure[1 orbit time, full SPH.] {\label{fig:SPH1orbit}\includegraphics[scale=0.4]{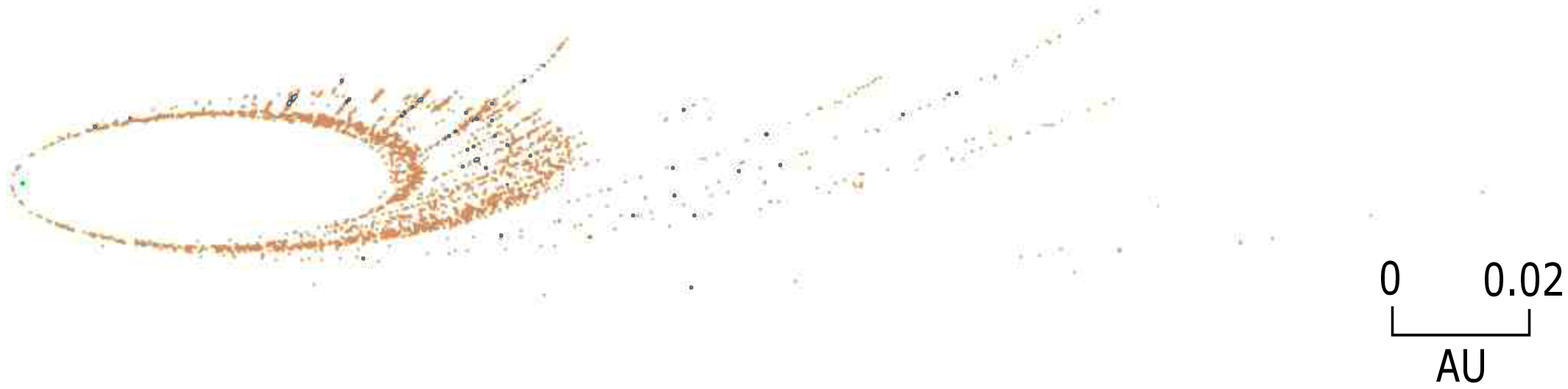}}
		\subfigure[1 orbit time, Hybrid.] {\label{fig:Hybrid1orbit}\includegraphics[scale=0.4]{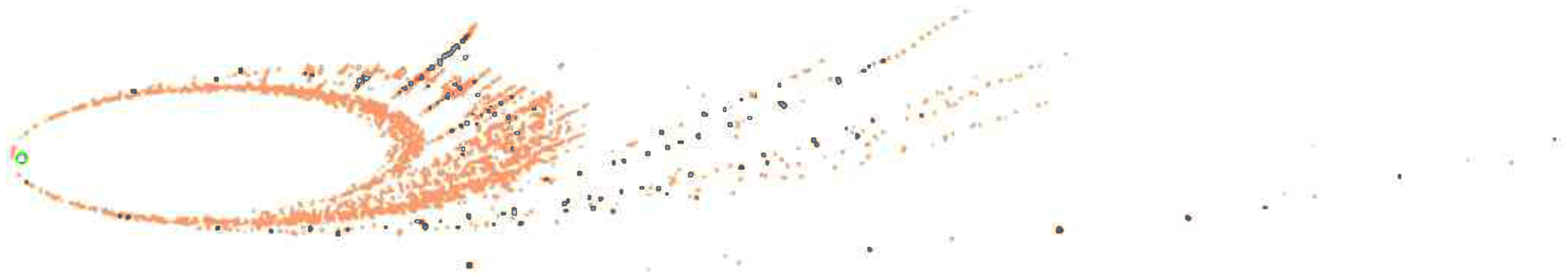}}
		\subfigure[2 orbit times, full SPH.]
 		{\label{fig:SPH2orbit}\includegraphics[scale=0.4]{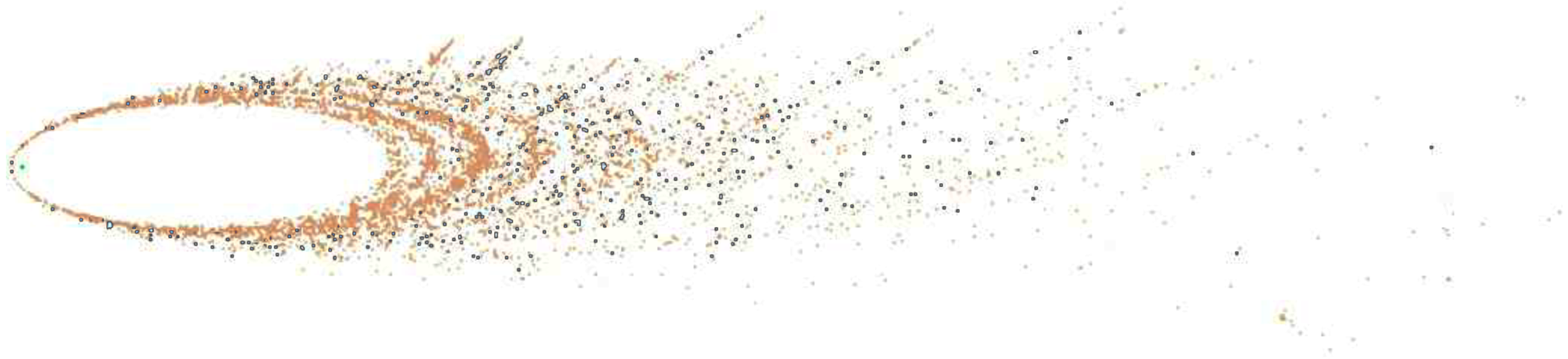}}
		\subfigure[2 orbit times, Hybrid.] {\label{fig:Hybrid2orbit}\includegraphics[scale=0.4]{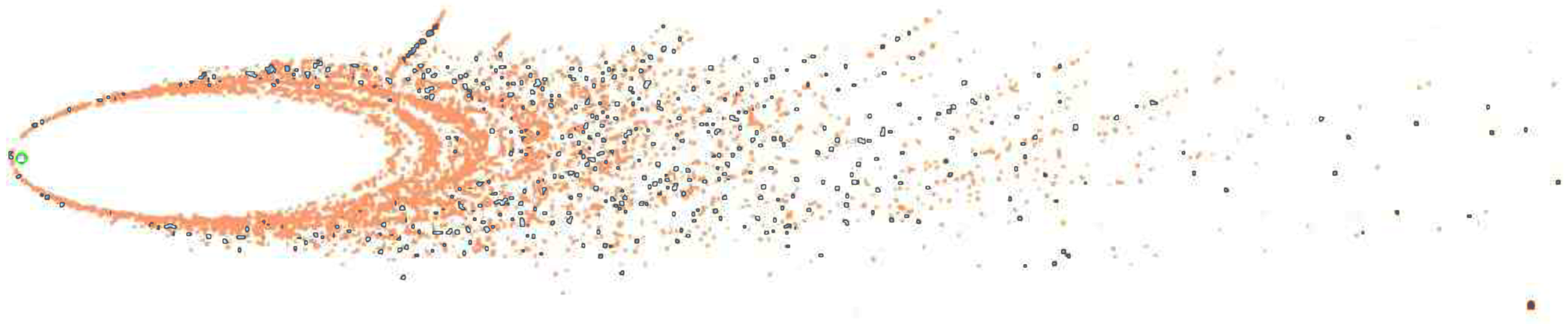}}
		\subfigure[9 orbit times, full SPH.] {\label{fig:SPH9orbit}\includegraphics[scale=0.4]{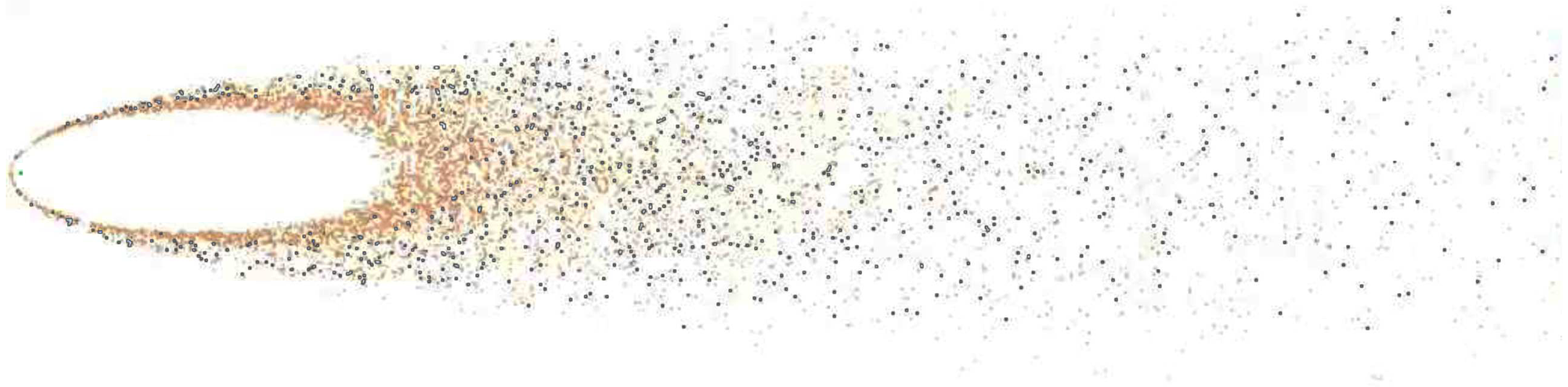}}
		\subfigure[9 orbit times, Hybrid.]
  		{\label{fig:Hybrid9orbit}\includegraphics[scale=0.4]{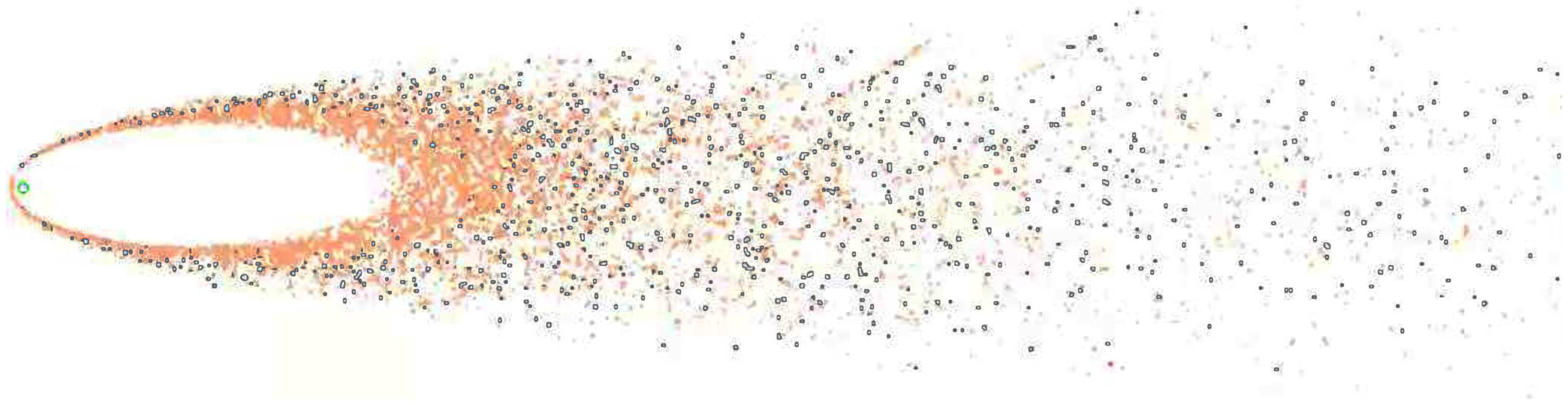}}		
	\end{center} 
	\caption{Top-down view of full SPH (left panels, dark tones) vs. hybrid method (right panels) discs. We compare the disc formation of a tidally disrupted Earth-sized planet ($a=0.1$ AU and $q=0.5R_{\odot}$) around a 0.6$M_\odot$ WD. Colour denotes composition: orange - rock ; black - iron; green - WD. The figure shows the disc formation progress in units of the original planet's orbital period (14.2 days). The hybrid model is generally indistinguishable from the full SPH simulation, but has a runtime of merely 2 days, instead of 100 days.}
	\label{fig:SPHVsHybrid}
\end{figure*}

Figure \ref{fig:SPHVsHybrid} shows that the hybrid model has fulfilled its goal. In this example we Compare the formation of a debris disc for a tidally disrupted Earth-sized planet around a 0.6$M_\odot$ WD. The planet's parameters are identical to those presented in Figure \ref{fig:zoomin0_5} with a pericentre distance of $q=0.5R_{\odot}$, and semi-major axis $a=0.1$ AU - an unrealistic and arbitrary choice, however one which makes the full SPH simulation runtime plausible (approximately 100 days). The left and right columns show the progress of full SPH and hybrid simulations, respectively. The colour scheme denotes composition: orange - rock ; black - iron; green - WD. The resolution is 10K particles. The formation progress is given in units of the original planet's orbital period, 14.9 days, and plotted for 1, 2 and 9 orbit times.

Note that the fragments (and star) in the hybrid simulation are scaled (magnified) with a factor of 1:100, which is sufficient for observing the angular size of the biggest among them. E.g., in the bottom of Panel \ref{fig:Hybrid1orbit}, we can see several fragments returning (clockwise) for a second disruption. Due to the magnification, these fragments are noticeably larger, being major, multiple-particle chunks from the original disrupted planet.

The final hybrid disc is almost indistinguishable from the full SPH simulation, yet it has a runtime of merely 2 days, instead of 100 days. In this example we had a factor 50 improvement in performance. However, by increasing either the resolution or the semi-major axis, the hybrid model would outperform the pure SPH simulation by a much higher factor.
 
Consider first the semi-major axis in this example. If we increase it by a factor of 100 to a more realistic value of 10 AU, the full SPH simulation runtime would (to first-order approximation) scale with the orbital period, thus taking about $100^{3/2}=1000$ times longer. On the other hand, raising it to 10 AU would make essentially no difference whatsoever in the hybrid model (the only departure from complete identity arises from changing the disruption regime, as suggested in Table \ref{tab:regimes}). The computation thus becomes plausible only with the hybrid model.

Now consider the resolution. Broadly speaking, SPH runtime scales like the resolution squared, so increasing the simulation from 10K particles to 500K particles would lengthen the runtime by a factor of 2500. In the hybrid model, the actual disruptions are also modelled with SPH (which means that the same rule applies), however the resolution is not constant. The runtime during the first flyby is equivalent to full SPH, since the planet initially contains all the SPH particles, however subsequent fragments become smaller and smaller, until eventually they may stop disrupting entirely. Hence, the simulation progress becomes exponentially faster. When $q$ is large, we have fewer yet larger (with more SPH particles) fragments, whereas deeper disruptions result in a multitude of smaller fragments (e.g., see Section \ref{SS:PericentreDependence} wherein extremely deep disruptions break the planet almost entirely into its constituent, SPH particles). In the latter case, the hybrid model will have substantially more fragments to iterate on, however they will cease disrupting quickly, typically by promptly reaching the minimum, single SPH particle size. In the former case, the hybrid model will have much fewer fragments to iterate on, however they may require several orbits until they all cease disrupting. 

We generally observe that the hybrid model runtime is the shortest for $q=0.1$. Simulations with higher $q$ values have a longer runtime by a factor of 2-4, however we see no clear proportion or relation between the runtime and $q$, since we suspect a more complex dependency, affected by more parameters than merely $q$. First, the runtime anti-correlates with the semi-major axis, since more eccentric disruptions decrease $r_\mathrm{crit}$ and produce less bound debris. Second, the runtime can either correlate or anti-correlate with the mass. On the one hand, increasing the mass gives a smaller relative fraction of bound material, and also increases the minimum fragment size (that of a single SPH particle), hence the simulation might be (depending on $q$) discontinued earlier for numerical limitations rather than for any physical reasoning. On the other hand, increasing the mass enlarges the dispersion in $\acute{a}$ (since $r_\mathrm{crit}$ moves in the opposite direction of the \emph{non-disruptive regime}, see Table \ref{tab:regimes}), so it can prolong the simulation. Overall, we emphasize that this factor is, at most, of the order of unity, hence the hybrid model performs well under all circumstances and for any combination of parameters.

In stark contrast, full-scale high resolution SPH simulations can be tracked within reasonable runtimes, \emph{only} when the disruptions are extremely deep (lowest $q$). The latter can lead to complete breakup (to the single SPH particle level) and a gravitationally unconfined stream, and then most of the bound mass falls back shortly after the first tidal disruption of the planet, and the complete absence of large fragments does not necessitate repeated disruptions. The implications are immediately evident - high resolution full SPH simulations will always remain computationally restricted to a very narrow portion of the disruption phase space.

We conclude that the hybrid model has fulfilled its aim. It produces similar results, yet enables unlimited increase to the semi-major axis, nullifying the constrains that have limited past simulations, and it also enables a huge increase in the resolution, particularly for large pericentre distances. To demonstrate its power in a more realistic scenario, we refer to Paper II, where we use the hybrid model to simulate a tidal disruption (originating at $a=150$ AU) at a resolution of 500K particles. Had we attempted to simulate the same scenario only with full SPH, the runtime could be estimated from the aforementioned arguments, as 100 days (see the simulation in Section \ref{SS:DiscFormation} which has the same $q$ and planet mass), times the increase in $a$ ($\times 1500^{1.5}$), times the increase in resolution ($\times 50^2$), or $\sim 1.45 \times 10^{10}$ days in total. Our hybrid simulation accomplished the same task in merely 40 days. We also note that the full SPH simulation from Section \ref{SS:DiscFormation} ran for 10 orbit times (of the original planet), whereas the hybrid model ran for 953 orbit times, when it reached the point in which the last fragment ceased from disrupting. Hence, not only is it more efficient, but also more complete, noting however that full completion is not necessarily a very significant criterion, since over 99\% of the disruptions occurred within the first few orbits anyway (e.g. see Section \ref{SS:DiscFormation}).

We further validate the hybrid model against the only previous work which studies disc formation through tidal disruption around WDs, performed by \cite{VerasEtAl-2014}. This work considers a $\sim$3 km asteroid, and so we generate a similar setup. We use the same orbit ($a=0.2$ AU), similar pericentre distance ($q=0.1 R_{\odot}$) and the same resolution (5K particles). The results are in very good qualitative agreement, producing a similar outcome to their Figure 10. We indeed form a ring of debris, as was our expectation from the \cite{VerasEtAl-2014} study, in addition to our theoretical predictions in Section \ref{S:Analytical}.


 \section{Future modifications and applications}\label{S:Future}
The new hybrid approach enables us to study a wide variety of problems, which are difficult to study using existing approaches. In Paper II we utilize the current code to perform a suite of hybrid simulations considering disc formation for a wide range of dwarf and terrestrial planets disruptions, pericentre distances and semi-major axes between 3 AU and 150 AU. However, we suggest several directions in which the current code can be further improved, and used for other general applications, as follows. 
 
\subsection{Disc formation by water-bearing objects}\label{SS:Water-bearing}
In this study we consider solely objects whose compositions are terrestrial-like, consisting of a rocky envelop and an iron core. However, recent observations \citep{FarihiEtAl-2013,RaddiEtAl-2015,XuEtAl-2017} in addition to our theoretical studies on water-bearing minor and dwarf planets around WDs \citep{MalamudPerets-2016,MalamudPerets-2017a,MalamudPerets-2017b} strongly reaffirm previous theoretical works \citep{SternEtAl-1990,JuraXu-2010}, and suggest that if such objects are common around main sequence stars, they should also be common around WDs. Our detailed thermo-physical simulations showed that much of their internal water content can be retained while their host stars evolve through the main-sequence, RGB and AGB high luminosity phases. It depends on an intricate set of parameters, including the host star's mass, in addition to their own size, composition, orbit and radionuclide abundances.

Water-bearing objects are generally both smaller and less dense, and therefore can disrupt within an even larger tidal sphere. What then might we expect from such tidal disruptions, and in what way would they differ from the terrestrial-like objects studied here?

As a general rule, the disruption modes and resulting discs (e.g. the semi-major-axes and size distribution) could be very similar to dry planetesimals, however the debris discs might form and evolve through a completely different route, according to the following arguments. Since the irradiation is proportional to the square of the orbital distance, disrupted water-bearing fragments (whose pericentre distances are between $\sim 0.1-1 R_{\odot}$) receive $\sim 10^4-10^6$ times the intrinsic luminosity of the WD during close approach, when compared to typical Solar system comets at 1 AU. Depending on its cooling age, WD luminosity ranges between $10^3-10^{-5} L_\odot$, hence we might typically expect $10^{-1}-10^9$ the amount of insolation, compared to Solar system comets (see \cite{MalamudPerets-2016} for discussion). However, we recall that in tidal disruptions the characteristic time a fragment spends near perihelion is only $\sim 10^{-1}$ days. Depending on the precise size of the fragment (which in turn might depend exactly on how deep the disruption is), we might expect various degrees of water sublimation rates.

It should also be important if the original object is homogeneously mixed or differentiated into a rocky core and icy mantle. For a differentiated object, fragments are expected to be composed of either one material or the other, and might have some cohesive strength. While the icy fragments will experience sublimation, which might decrease their size between disruptions, the rocky fragments would evolve in much the same way as described in this study, since refractory materials have much higher sublimation temperatures \citep{RafikovGarmilla-2012,XuEtAl-2018}. For a homogeneous object, tidally disrupted fragments are rather expected to remain homogenous, and since the original object is likely small (or else it would differentiate), the fragments would be even smaller - most likely rubble piles dominated by gravity alone. Outgassing volatiles might carry with them dust or pebble-like silicate grains, and then the two distinct compositions might evolve on very different timescales. For virtually any WD (with $L<0.1L_\odot$) the radiation forces are too feeble compared to the gravitational forces and are thus unable to disperse the gas \citep{BonsorWyatt-2010,DongEtAl-2010,VerasEtAl-2015}, which most likely would accrete onto the WD by experiencing ionization and then being subject to the magneto-rotational instability \citep{KingEtAl-2007,Farihi-2016}. Whereas the silicate grains are slower to evolve and might be subject to PR drag \citep{VerasEtAl-2015} and collisions \citep{KenyonBromley-2017,KenyonBromley-2018}.

These, as well as other considerations, suggest a level of complexity that needs to be addressed in future dedicated studies. In particular, it might require an approach which utilizes the hybrid technique in combination with other numerical methods or analytical calculations. E.g., the sublimation evolution of fragments may be studied either through complex numerical simulations like the ones used by \cite{MalamudPerets-2016,MalamudPerets-2017a,MalamudPerets-2017b} or via detailed analytical treatment \citep{BrownEtAl-2017}.

\subsection{Inclusion of strength/porosity models}\label{SS:Strength}
The original developers of SPH considered the dynamics of fluid flow, governed by a set of conservation equations (mass, momentum and energy). An equation of state completes the scheme by relating the various thermo-dynamical variables. As the SPH technique evolved, more advanced models have emerged which incorporated representations for elastic solids, elasto-plastic solids, fracture/damage in solids and inclusion of sub-resolution porosity in small objects. All of the aforementioned models are in fact implemented in the SPH code \emph{miluphCUDA}, yet they are not used in this study. In this study we perform only fully hydrodynamical SPH simulations, neglecting material strength.

For impact collision modelling, it is a well-established fact that material strength can greatly affect simulation outcomes. See e.g. some recent papers by \cite{BurgerSchafer-2017,GolabekEtAl-2018} which are relevant to the size range investigated in our study. For disc formation by tidal disruptions, however, no previous work has ever been performed, to the best of our knowledge, that methodically investigated SPH simulation outcomes when comparing various strength models. It is a well known fact that for very large objects (in the range of hundreds to thousands of km, depending on the composition), self-gravity dominates in determining the size of the tidal sphere, as it opposes the tidal force (see Section \ref{SS:DiscFormation}). However, for smaller bodies, the internal material strength takes over as the dominant force \citep{BrownEtAl-2017}. Therefore, there is a size range in which material strength can be rather important for tidal disruptions and their outcomes.

It is however incorrect that material strength always becomes more important with diminishing size. There exists a class of small objects, in the range of hundreds of meters to a few km, for which the effective strength to resist global deformation is once again low, when it is controlled by fractures or flaws \citep{Jutzi-2015}. These so called rubble-piles are dominated by gravity, and they may easily break apart by failure in their low-strength internal fault surfaces. Past models \citep{BenzAsphaug-1994} have often treated such fractured material as completely strength-less, with pressure independent yield criterion, whereas some newer constituent models consider internal friction and pressure-dependent yield criterion \citep{CollinsEtAl-2004,Jutzi-2015} since it is known that the shear strength of rocks is pressure-dependent. According to \cite{BrownEtAl-2017}, the fragments that emerge from tidal disruptions of fractured rubble pile structures, may potentially be considered as monolithic objects if at some smaller scale they once again start to have internal cohesion, although what this scale might be is not yet certain. If we rely on the cohesionless asteroid spin-barrier as evidence, it might be around 150-300 m \citep{PravecEtAl-2002}.

We should consider the possibility that initially cohesive bodies may lose some of their strength when undergoing partial disruptions, or perhaps during full disruptions which produce recycled, fully-damaged, rubble-pile fragments. Then tidally disrupted rubble-piles might once more inject smaller particles with some internal cohesion. All of these aspects require detailed research and merit future work in these directions.

\subsection{Stream gravitational confinement: fragmentation and intra-collisions}\label{SS:Fragmentation}
In Section \ref{SS:PericentreDependence} we showed that tidal streams may or may not be gravitationally self-confined, and fragment under their own self-gravity. A preliminary analysis was performed, equating theoretical predictions and calculations with numerical simulations. We have shown that in some cases the stream undergoes fragmentation, with very good agreement between simulations and theory, while extremely deep disruptions seem to inhibit fragmentation.

This result was interpreted as being linked to the stream's free-fall timescale through the $\alpha$ parameter, which appeared to be strongly dependent on the breakup distance. It is an important behaviour which necessitates further investigation, and must rely on a much more extensive grid than was possible in this paper, which had a broader goal and focus. For future applications of the hybrid model, we should numerically determine $\alpha$ by exploring a wider parameter space of breakup distance and planet size. We would then be able to semi-analytically determine the precise SPH duration required in each application.

Additionally, we have shown that fragments, shortly after being formed, collide among themselves, prior to reaching a more stable, longer-term collisionless state. In our hybrid simulations we have largely neglected the importance of this stage. However, complete understanding of the tidal disruption and disc formation phenomena entails some further development of the theoretical framework governing this process, including a detailed model and an investigation of the collision outcomes / the production of second-generation fragments. Particularly, the size distribution and abundance of collision-induced small particles is an important question, since swarms of small particles and dust can manifest as strong transit events, while larger fragments cannot. For this purpose, we could further examine such collisions in high resolution, using typical impact parameters from our existing simulations.

\subsection{Disc circularization and evolution}\label{SS:CircularizationEvolution}
Our paper and the \cite{VerasEtAl-2014} study, deal with the initial formation phase of a disc, triggered after a tidal disruption following a close approach to a WD. While both studies result in completely different debris discs, the outcomes nevertheless share a common morphological characteristic -- eccentricity. The \cite{VerasEtAl-2014} study considers an asteroid, which, after disruption, forms a narrow eccentric ring of particles following the original asteroid trajectory. Typically, the asteroid must originate from a distance of at least a few AU. As such, the asteroid and resulting ring have an eccentricity approaching 1. Our study considers much bigger objects, up to terrestrial planet sized, and originating from various potential regions of a planetary system, up to hundreds of AU. When disrupted, they form dispersed discs of interlaced elliptic eccentric annuli, extending from as little as $\sim$0.05 AU (in the most extreme case) to well beyond the original planet orbit (see Figure 1 in Paper II). The corresponding eccentricities of fragments in such a debris disc are therefore at the minimum 0.9 and typically much more.

Taking a leap forward in time, the studies of \cite{KenyonBromley-2017,KenyonBromley-2018} focus rather on the final formation sequence of the disc, when it reaches a much more compact state (the eccentricity of order $\sim 0.01$). Here, collisional grinding of large particles rapidly pulverize them to mere dust and gas. 

We are \emph{missing} an important link in between those stages. \cite{VerasEtAl-2015} consider disc shrinkage through the drifting of small micron-to-cm sized particles by PR drag, however it is unclear that this particle size range constitutes a significant mass fraction of the disc. Rather, various arguments throughout this paper (and in particular Appendix \ref{A:Radiation}) emphasize the potential importance of the Yarkovsky effect, which applies to more sizeable fragments up to hundreds of m. The Yarkovsky effect could be an important agent for circularizing the disc, however it remains to be properly explored in this context \citep{VerasEtAl-2015}. Rotational statistics (presented in Paper II) could provide a key input for such models. Alternatively, collisional cascade might potentially break the fragments on longer evolutionary timescales, such that a more significant fraction of the disc could evolve through PR drag \citep{WyattEtAl-2011}. This possibility requires further consideration in the context of our study.

Another direction that may facilitate disc shrinkage has been recently proposed, from a slightly different angle. \cite{GrishinVeras-2019} suggest that small exo-Kuiper or exo-Oort like objects in the 0.1-10 km size range, can be captured by a compact gaseous disc in the vicinity of a WD. If so, the same analytical formalism may be used in order to study the fate of similar-sized fragments that form in our study, and which may likewise experience some gradual dissipation as they interact with such a gaseous component. Once again, the starting point for the calculation could be the disc layouts that emerge from our hybrid models.
In Section \ref{SS:PericentreDependence} we show that tidal streams may or may not be gravitationally self-confined, and fragment under their own self-gravity. A preliminary analysis is performed, equating theoretical predictions and calculations with the actual numerical simulations. We have shown that in some cases the stream undergoes fragmentation, with very good agreement between simulations and theory, while extremely deep disruptions seem to inhibit fragmentation.

This result is interpreted as being linked to the stream's free-fall timescale via the $\alpha$ parameter, which appears strongly dependent on the breakup distance. It is an important behaviour which necessitates further investigation, and must rely on a more extensive grid than was possible in this paper, which had a broader goal and focus. Given our results in Section \ref{SS:PericentreDependence}, we need to further explore the parameter space of breakup distance and planet size.

We advocate that this missing link should attract a much greater focus than it has thus far. For example, consider the disintegrating object around WD 1145+017 \citep{VanderburgEtAl-2015}. In \cite{VerasEtAl-2017}, a differentiated body initially placed at nearly circular orbit might steadily disrupt its mantle to produce a compatible signal with hour-scale periodicity, but how does it assume such an orbit in the first place? The ideas discussed in Paper II suggest that it might simply be in a more advanced evolutionary state. That evolution is our missing link.

\subsection{Ring formation around giant planets}
\label{SS:RingFormation}
Ring formation around giant planets has been previously associated with tidal disruptions, either by a primordial satellite after inward migration towards the planet \citep{Canup-2010} or by an occasional passing object \citep{Dones-1991,HyodoEtAl-2017}. These various scenarios were modelled via (fluid) SPH simulations.

In the latter case, the captured fragments from the disruption typically move on very eccentric orbits. Hence, \cite{HyodoEtAl-2017} stop their SPH simulations (using a resolution of $10^5$ particles) after the initial flyby, and in any case prior to the fragments reaching their apocentres. They use follow-up N-body simulations for the longer-term evolution of fragments, including the effect of a planet’s oblate potential. However, the streams that emerge in their SPH simulations are gravitationally self-confined and large fragments are bound to return for subsequent disruptions and should be further modelled with SPH to obtain an accurate particle size and orbital distributions.

We propose that the hybrid model is suitable to this task. It can follow the tidal disruption of passing objects at superior resolution and track further disruptions of returning fragments. Subsequent N-body simulations can be employed in a similar manner to \cite{HyodoEtAl-2017} or Appendix \ref{A:Collisionless}, however only at later times. Furthermore, internal strength treatment, as suggested in Section \ref{SS:Strength}, might prove to be very important. Likewise, modeling small passing objects as porous might be important, since their sound speed is typically one or two orders of magnitude lower compared to consolidated material, thus fractures can propagate throughout the body much more slowly, with a timescale comparable to the encounter timescale (in contrast, see \cite{Dones-1991}).
	
\subsection{Formation of families of Sun-grazing comets}\label{SS:sun-grazing}
The tidal disruption of minor planets which approach close enough to the sun to be disassociated by tidal forces, is a relatively little studied aspect of their fate, when considering detailed numerical approaches. As mentioned in Section \ref{S:Intro}, the only previous numerical study was performed by \cite{WeissmanEtAl-2012}, based on the n-body code developed by \cite{MovshovitzEtAl-2012}.

Using our hybrid model for tidal disruptions, we have a way of studying tidal disruptions for any progenitor orbit. We can track their repeated passages around the Sun with unprecedented resolution, and try to explain their emerging enigmatic observational attributes \citep{GranvikEtAl-2016}. From such a study we can potentially infer their characteristic properties.

\section{Summary}\label{S:Summary}
Our study introduces a new method for performing high-resolution, tidal disruption simulations, at arbitrary orbits. We call the technique the hybrid method, since it combines full SPH simulations in the relevant tidal disruption spatial domain, but also treats the orbit of tidally disrupted fragments analytically by instantaneously transferring them back to the tidal sphere for subsequent disruptions. Hence, the hybrid approach saves a tremendous amount of computational power, opening new possibilities for studying the long-term formation sequence of tidal disruption debris discs, which have not been possible with any existing model.

Prior to introducing our hybrid technique, however, we performed the following steps. First, we outlined a simple analytical impulse approximation model for treating single tidal disruptions. Using this model, we showed that tidal disruptions generally depend on a number of parameters. Nevertheless, given characteristic Solar system distances, small asteroids usually form a narrow ring in which all the asteroid material is bound to the star and remains on semi-major axes close to that of the original progenitor. Such disruptions were therefore termed the 'non-dispersive regime'. In contrast, larger dwarf or terrestrial planets usually form a completely different debris disc, in which just over half of the material becomes tightly bound to the star (compared to the progenitor orbit) and the other half becomes unbound, assuming hyperbolic trajectories. Such disruptions were termed the 'bi-modal regime'. 
	
While providing invaluable insights into the process of single tidal disruptions, the analytical model is however incorrect, since its assumptions are merely approximations. In reality: the disruptions are not instantaneous but gradual; the object is not spherical at breakup but elongated; the dissociation among the constituent particles is not complete unless the disruption is very deep; self-rotation prior to the disruption affects the disruption outcomes; and the model doesn't account for inhomogeneous objects consisting of various materials/densities/internal strengths etc. The most important point, however, is that unless tidal disruptions are extremely deep, they do not finish after merely a single passage through the tidal sphere. Large fragments can repeatedly return for subsequent disruptions, and given their new sizes and orbits they can disrupt in entirely different regimes compared to their original progenitor. All of these points merit the use of hydrodynamical simulations for a more realistic treatment of both single, as well as multiple repeated disruptions, which are required in order to form a disc of debris.
	
We begin our numerical study by performing full hyrdodynamic SPH simulations of various tidal disruptions. For the simulated bodies, low resolution and a small semi-major-axis are required in order to track the formation of the debris disc, otherwise the problem becomes computationally impossible. We perform the first ever tidal disruption simulations of terrestrial-sized planets inside the tidal sphere of a WD, at various pericentre distances. For large pericentre near the Roche limit, we show that the disruption is partial, shedding only a small fraction of the planet's mass, emanating from its outer portions. If the pericentre is halved, the disruption is full. It forms a classic narrow stream of debris, however the stream is gravitationally self-confined, and it collapses to form multiple fragments, which later also collide and merge among themselves, forming smaller second-generation particles. An analytic fragmentation timescale model is introduced, and we show that it compares well with the measured fragmentation timescale. If the pericentre is a small fraction of the Roche limit, the disruption is both full and violent, such that the stream is gravitationally unconfined and the planet becomes almost entirely disassociated into its constituent particles. We discuss the possibility that interstellar asteroid Oumuamua is a hyperbolic fragment from a tidal disruption around a WD, formed by either one of the two \emph{full} disruption modes.
	
We then show a full SPH simulation of debris disc formation from start to finish, for a disrupted planet passing within but near the Roche limit. Even for this limiting case, the disc evolves rather quickly. After a few orbital times of the original planet, all fragments finish their repetitive sequence of tidal disruptions. The dispersed structure of the disc is made of a superposition of interlaced elliptic eccentric annuli, that form consecutively as time passes. An emerging feature in this evolution is the tidal spin-up of the original planet and other fragments. Their excited self-rotation rates are rapid, typically corresponding to a large breakup velocity fraction. This fast rotation in the prograde sense significantly expedites the disruption processes in all but the first flyby, since the original planet typically rotates very slowly compared to its breakup velocity. We nevertheless also quantify the effect of an initially rotating planet, and find it to be neither significant nor entirely negligible.
	
Afterforwards we discuss the new hybrid concept. Each fragment disruption is modelled individually around the star. The relevant spatial domain is the Roche limit, where tidal forces dominate, in addition to the fragmentation domain. Subsequently, each fragment orbit is analysed and its next position for the successive tidal disruption is calculated. Our hybrid method entails a couple of assumptions, namely, that the fragments are largely unaffected by radiation effects and that the disc is collisionless. We review the literature to show that radiation is not very important during the typical disc formation timescales. We come to a similar conclusion about collisionality, by handing over fully developed SPH discs to a newly modified N-body code, and quantifying the amount of collisions. Not only are collisions scarce (to the 1\% level) in the timescale relevant to formation, but they are also restricted to a spatial domain inside or near the Roche limit. The same conclusion is supported for higher resolutions by analytic arguments.

We then introduce our code specifics, testing the outcomes of the hybrid model against identical full-scale SPH simulations. We conclude that the hybrid model produces essentially identical results, while computationally outperforming the full SPH simulations by many orders of magnitude. It enables unlimited increase to the semi-major axis, nullifying the constrains that have limited past simulations, and at the same time also enables a huge resolution increase. 

We conclude the paper in Section \ref{S:Future} by listing a number of important future directions and applications which could further improve the hybrid model. To name the most important ones, the SPH code could also utilize more advanced internal strength, porosity and damage models, for much more realistic tidal disruption outcomes. We could also extend the model to be used in modeling water-bearing bodies in addition to dry ones. We suggest a number of emerging applications which could directly benefit from the hybrid model, and which cannot be numerically modelled by other methods due to computational limitations.

\section{Acknowledgment}\label{S:Acknowledgment}
We wish to thank the anonymous reviewer for excellent suggestions and comments that have greatly improved this manuscript. UM and HBP acknowledge support from the Minerva center for life under extreme planetary conditions, the Israeli Science and Technology ministry Ilan Ramon grant and the ISF I-CORE grant 1829/12. HBP is supported by the Kingsley distinguished visitor program in Caltech.

\newpage
	
\bibliographystyle{mnras} 
\bibliography{bibfile}     

\newpage
\begin{appendices}

\section{Initial planet rotation}\label{A:RotationDependence}
As we have shown Section \ref{SS:DiscFormation}, self-rotation can be rather important in determining the outcome of tidal disruptions. We should nevertheless make the distinction between returning fragments, whose self-rotation can be significantly excited during previous disruptions, and the original disrupting planets which rotate slowly (in the Solar system typical rotation periods are of the order of a few $10^1$ h). We have shown that when the breakup velocity fraction is very small ($\lambda\ll 1$), self-rotation should have a negligible effect on the tidal force. 

However, even a slow rotation may have a noticeable effect, causing the material to fall back faster (for prograde rotation) or slower (for retrograde rotation), by modifying the minimum apocentre distance in the bound tidal stream. This modification occurs since the star has to first spin down a retrograde spinning planet, delaying the disruption. Without repeating their calculations, we base the following formulations on the study of \cite{GolightlyEtAl-2019a}. It considers the related phenomena of stellar tidal disruption events (TDEs), in which stars are tidally disrupted around super-massive black-holes. It is typically assumed that $\lambda\ll 1$ because such stars usually rotate slowly, unless one considers stars in the so-called loss-cone diffusive regime \citep{StoneMetzger-2016}, which are gradually pushed to smaller orbits and their rotation can spin up by the tidal field before the disruption occurs \citep{AlexanderKumar-2001,AlexanderKumar-2002}. \cite{GolightlyEtAl-2019a} analysed the effect of initial stellar rotation on TDEs, using analytical calculations and numerical models with moderate $\lambda$ up to 0.2. They show that (given the \emph{bi-modal disruption regime} where $r_\mathrm{crit}\ll R$ and the disruption is full) for the most-bound fragment (or gas parcel, in their case) in the inner stream, we can re-write the semi-major axis from Table \ref{tab:regimes} such that:

\begin{equation}
\acute{a}=-\frac{d^2}{2R(1\pm \sqrt{2} \lambda)}
\label{eq:Semi-majorAxisRotationCorr}
\end{equation}

The positive sign applies to prograde rotation whereas the negative sign applies to retrograde rotation. When $\lambda=0$, the correction term goes to 1, and Equation \ref{eq:Semi-majorAxisRotationCorr} recovers the former $\acute{a}$ from Table \ref{tab:regimes}.

For the most-bound fragment, the first to return to the star, we shall note two additional equations. Its energy is given by:

\begin{equation}
E=-\frac{GM}{2\acute{a}} = -\frac{GMR}{d^2} \left( 1 \pm \sqrt{2}\lambda \right)
\label{eq:EnergyRotationCorr}
\end{equation} 

Also, the fragment's return time is given by Kepler's law. It reads:

\begin{equation}
T=\frac{2\pi}{\sqrt{GM}} \acute{a}^{3/2} = \frac{\pi d^3}{\sqrt{2GM}} \left( R( 1 \pm \sqrt{2}\lambda )\right)^{-3/2}
\label{eq:ReturnTimeRotationCorr}
\end{equation} 

A possible way to validate this equation is to compare the SPH outcomes of prograde versus retrograde rotating planets. The ratio between their minimum return times should accordingly be $((1 + \sqrt{2}\lambda)/(1 - \sqrt{2}\lambda))^{-3/2}$, and can be compared with detailed SPH simulation results.

For this comparison we perform additional SPH simulations. We take a typical rotation period of 20 h for the planet and a pericentre distance of 0.5$R_{\odot}$ (because it yields a fragmented, gravitationally-confined stream, following a full disruption, as we have seen in Section \ref{SS:PericentreDependence}). Since in these simulations we are not interested in the full formation of the disc, but only the initial properties of the stream after the first tidal approach, we are less restricted by computational time limitations and thus our semi-major axis is not constrained like in Section \ref{SS:PericentreDependence}, giving us freedom to take realistic values, which will also bring us closer to the \emph{bi-modal disruption regime}. We use $a=3$ AU, as motivated in Section \ref{SS:SPHOutline}.

\begin{figure}
	\begin{center}	
		\includegraphics[scale=0.475]{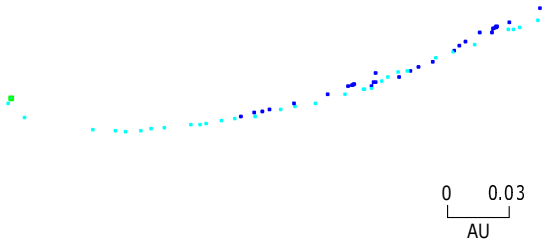}
		\caption{First bound fragments returning to Pericentre, for two Earth-sized planets with $a=3$ AU and $q=0.5 R_{\odot}$, initially rotating in the prograde (blue pixels) and retrograde (purple pixels) sense, with a 20 h rotation period. The elapsed time from the initial disruption around a 0.6$M_\odot$ WD (green pixel) is $t=7.64$ days. The prograde initial rotation results in a bigger spread in orbital energies.}
		\label{fig:ProgradeVsRetrograde}
	\end{center}	
\end{figure}

Figure \ref{fig:ProgradeVsRetrograde} shows the vicinity of the WD (up to a distance of about $\sim$0.3 AU from the star). The image is composed of two, superimposed disruption images, one for a prograde-rotating planet (whose fragments are shown in red) and another for a retrograde-rotating planet (fragments in yellow). It captures the first returning fragments from these disruptions. As can be seen, by $t=7.64$ days, the most bound fragment from the disrupted prograde-rotating planet has reached the pericentre distance of $q=0.5 R_{\odot}$. The most bound fragment from the retrograde-rotating planet still lies about 0.12 AU away, and will only reach pericentre at $t=9.88$ days. The ratio of these two return times equals 0.773. As we have shown, using Equation \ref{eq:ReturnTimeRotationCorr} we can extract $\lambda$, given this ratio, and in doing so we obtain a value of 0.06. This value is in good agreement with that used in our simulation, of $\lambda \cong 1.4/20 \sim 0.07$. Hence, our simulations are shown to agree, to zeroth order, with the analytical derivations. The small discrepancy most likely arises from various simplifying assumptions made in the analytical model.

How does the early or late arrival of returning fragments changes the formation of the disc? We argue that since slow rotation does not really change the geometry of the stream after its initial formation, but merely its length, the overall formation of the disc is not affected much. Rotation would simply offset the innermost ring of material in the disc (either inwards or outwards, for prograde and retrograde assumed rotations respectively), but the overall shape and properties of the disc would remain mostly unchanged. Our conclusion from this analysis is that one may therefore typically ignore the initial rotation of the planet, since it generally constitutes a second-order effect only. Throughout this paper, we assume that all planets are initially non-rotating, thereby dismissing the need for an additional free parameter in our suite of simulations, however with the caveat that even slow initial rotation is not completely negligible. 

We further note that inclined rotation could provide dispersion in the vertical axis, potentially increasing the disc height. This effect should merit further investigation in the future.

\section{Radiation effects?}\label{A:Radiation}
We examine the assumption that radiation effects can be neglected. Our conclusions are largely based on the study of \cite{VerasEtAl-2015}, which considers how the radiation from the WD can shrink an initially extremely eccentric ring of debris. To our knowledge, this is the only study which attempted to fundamentally address this important problem. In the predecessor \cite{VerasEtAl-2014} study, the same authors show that a ring of debris may form through the tidal disruption of an asteroid. The disruption is assumed to be very deep, with $q$ on the order of 0.135-0.27 $R_{\odot}$, and the asteroid is assumed to be a rubble pile. It is suggested that this particular combination results in very small debris, from micron to centimetre sized. 

\cite{VerasEtAl-2015} show that due to the effects of Poynting–Robertson (PR) drag, the shrinking time of the disc to a completely circularized one could be anything from years up to hundreds of millions of years, depending on the exact size of the debris and particularly the cooling age of the WD, which determines its luminosity. They note however that in cases where the debris are larger than centimetre-sized, PR drag becomes ineffective, and instead the seasonal Yarkovsky effect dominates. Estimating the precise drift contribution of the Yarkovsky effect is nevertheless a challenging task, especially when the affected debris are on an extremely eccentric orbit \citep{VerasEtAl-2015}, and remains to be explored further. Objects larger than typical asteroids are not sensitive to the Yarkovsky effect, however such sizes are never considered in the \cite{VerasEtAl-2015} study, since they only discuss small asteroids to begin with.

In our study, we do not merely consider small asteroids, which may or may not be actual rubble piles, but instead large dwarf or terrestrial sized planets. We show in Section \ref{SS:PericentreDependence} that tidal disruptions heavily depend on the exact pericentre distance, and may result in either partial or full disruptions. 

When the disruption is gravitationally unconfined, the debris may indeed be small, however determining how small is a complex and challenging problem \citep{Rafikov-2018}. When it is gravitationally self-confined, particles mostly coagulate into much larger fragments, regardless of the initial size distribution of the debris when they are still inside the tidal sphere. Initially, therefore, they might be orders of magnitude larger than the sizes relevant to either PR drag or Yarkovsky, however after repeated disruptions that occur during the formation of the disc, their sizes will gradually diminish (see also Paper II). Eventually they should reach some characteristic size which inhibits further breakup, depending on their intrinsic internal strength \citep{BrownEtAl-2017}, in addition to the pericentre distance (see discussion in Section \ref{SS:Strength}). It is essentially unknown what their size might be at this stage, although according to \citep{PravecEtAl-2002} or \cite{BrownEtAl-2017} the likely size could be in the range of tens of meters to hundreds of meters.

Let us first look at the importance of PR drag. Given the statements in the last paragraph, it is somewhat unlikely that most of the debris consist of micron-to-centimetre sized particles. As we shall see in Appendix \ref{A:Collisionless}, perhaps during later times, particle sizes might diminish by collisional grinding (see also \cite{WyattEtAl-2011}), however this process is not important during the \emph{initial} formation stage of the disc, which is relevant to our hybrid model, and in turn, the same can be said for PR drag. Additionally, we recall that small particles may arise from the short collisional phase, in which fragment mergers produce a secondary cloud of debris (see Section \ref{SS:PericentreDependence}). In this case too, the mass fraction of micron-to-centimetre particles in the disc is probably not significant, in addition to the fact that they lie well within the gravitational potential of their larger parent fragments, and thus the dominant forces are gravitational, not those from PR.

Let us now discuss the Yakovsky effect. As previously stated, often our simulations involve very large fragments. In fact, the smallest fragment size we may obtain is that of a single SPH particle. Given resolutions that range from 10K to 500K particles, and typical masses of planets/dwarf-planets, SPH particle radii range between 10-250 km. The seasonal Yarkovsky effect is important for decimetre to 10 km sized particles, thus resolution limitations make it impossible for us to self-consistently correct for Yarkovsky. 

Laying aside numerical limitations, however, it is clear that the Yarkovsky effect could be important for many discs. Improved resolution and utilizing creative approaches in the future (or simply modeling small progenitors) may allow us to reach the relevant fragment sizes. Determining the exact rotation rate of each fragment, it should therefore be possible to improve on our simple analytical Keplerian treatment by adding a correction-term, accounting for the Yarkovsky effect. In the meantime, as suggested by \citep{VerasEtAl-2015}, more progress is required in order to develop a better theoretical understanding of the seasonal Yarkovsky orbital shrinking effect in highly eccentric orbits. This development is however left to future studies, exceeding the goals of this paper, and could benefit from recent publications such as that of \cite{VerasEtAl-2019}.

We conclude that radiation effects are often inapplicable to our simulations. Even in cases where they should be considered, like in the Yarkovsky effect, we emphasize that the hybrid method is meant only to solve the problem of initial disc formation. The results from Section \ref{S:FullSPH} (as well as Paper II) place a tight limit on the formation timescale of the disc, such that they often finish forming within only a few orbits (of the original planet). We therefore argue that radiation effects are to be considered mainly during the longer-term circularization/shrinking phase of the disc (see discussion in \ref{SS:CircularizationEvolution}), and not so much during its initial formation. Considering the analysis of \cite{VerasEtAl-2015}, this statement seems rather judicious for PR drag, and would merit some further investigation for the Yarkovsky effect. In any case, if radiation effects are found to be important for the hybrid method, they are certainly equally important for full-scale SPH simulations. In fact, the hybrid method might actually be superior in this regard, due to the possibility of Yarkovsky-correcting its analytical Keplerian treatment.

\section{Collisionless disc?}\label{A:Collisionless}
We examine the assumption of a collisionless disc. We note that there are two different regions to consider. First, we consider the region close to the star, where we employ the SPH code to model the tidal disruptions. We recall that the SPH code simulates each disruption independently, as though the region near the star is devoid of other particles. In other words, the model assumes that the immediate region of the star is collisionless. Second, we consider the region far from the star, where our hybrid model instantaneously transports fragments back to the tidal sphere, as though that region is likewise, collisionless.

In Section \ref{SS:PericentreDependence} we have seen that the breakup and fragmentation phases are followed by a relatively short collisional phase, wherein some fragments in the stream may merge. Following this phase, however, collisions (or gravitational interactions) may only occur in two ways. (a) fragments may randomly collide with pre-existing materials, which happen to occupy the space in which they traverse; and (b) fragments may collide among themselves during the formation progress of the disc. Since repeated disruptions continually increase the number of fragments and their semi-major axes dispersion, the probability for collisions raises as a function of time.

For the purpose of this paper we shall assume that pre-existing material in orbit is sparse enough so that collisions of the former type can be neglected, especially since the initial formation of the disc is a relatively rapid process, and during this time they are unlikely to occur. We however suggest that \emph{after} the disc initially forms, it is perhaps more plausible for it to evolve on longer timescales by interacting with existing materials (e.g., a system of planets). This sort of interaction remains to be explored in future studies, although we generally find this possibility rarer than collisions of the latter type. The kind of orbits that typically emerge in our models with dispersed discs place the overwhelming majority of fragments in relatively tight orbits, at characteristic distances of up to a few AU. Planets surviving the post main-sequence stellar evolution are typically exterior to such orbits. Hence, unless they migrated inwards during the WD lifetime \citep{VerasFuller-2019a,VerasFuller-2019b}, direct interactions with a system of planets is not expected, while secular and resonant interactions with wider orbit planets are still possible, and their potential effects certainly imply that further studies might be required.

In order to determine if the disc is also collisionless among its own fragments, we employ the following approach. We would like to show that even a fully formed disc, let alone one in the process of formation, would be largely collisionless. We thus take a set of fully formed discs, with a small semi-major-axis of 0.1 AU, allowing us to carry them out to conclusion with full SPH simulations, until all fragments cease disrupting. These simulations are similar to the one presented in Section \ref{SS:DiscFormation}, with WD mass of 0.6 $M_{\odot}$. We consider three choices for the pericentre distance: 0.1 $R_{\odot}$, 0.5 $R_{\odot}$, and 1 $R_{\odot}$. Our results are obtained for a Mars-sized planet.

These fully formed discs are then handed over from SPH to the open source N-body simulation REBOUND, via a special tool which we have developed. This hand-off tool is also designed to keep a detailed record of all mergers, using the existing REBOUND reb\_collision\_resolve mechanism. We have modified the REBOUND source code to both record the mergers and their inherent attributes (time, position, masses and composition) in a separate set of files, in addition to keeping track of the relative compositions of merged particles (utilizing the 'additional properties' built-in formalism), which we then use in order to calculate a more realistic physical collision-radius for the REBOUND particles. We use the IAS15 integrator - a fast, adaptive, high-order integrator for gravitational dynamics, accurate to machine precision over a billion orbits \citep{ReinSpiegel-2015}. Our implementation also utilizes openmp, to get about a 30\% improvement in runtime when using 8 cores in parallel (more cores gain no further improvement).

\begin{figure}	
	\begin{center}		
		\subfigure[$q=1 R_{\odot}$.] {\label{fig:collisions1q}\includegraphics[scale=0.475]{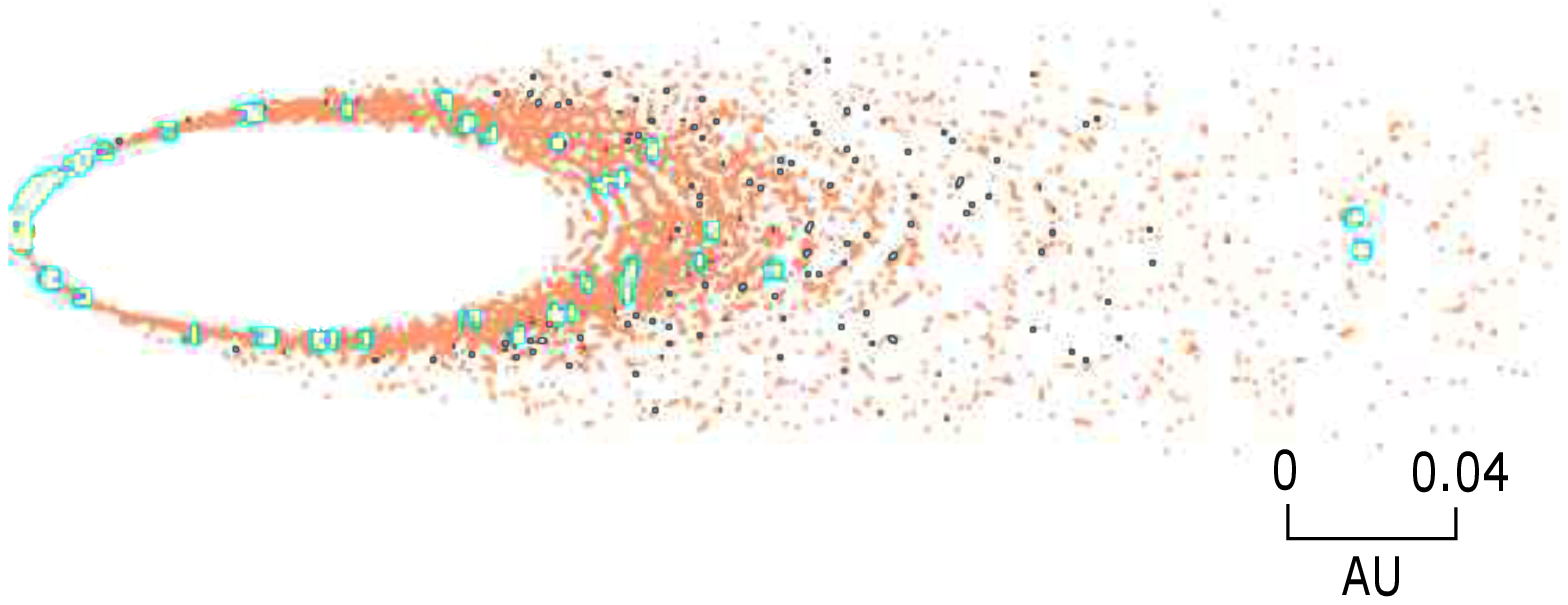}}
		\subfigure[$q=0.5 R_{\odot}$.] {\label{fig:collisions0_5q}\includegraphics[scale=0.475]{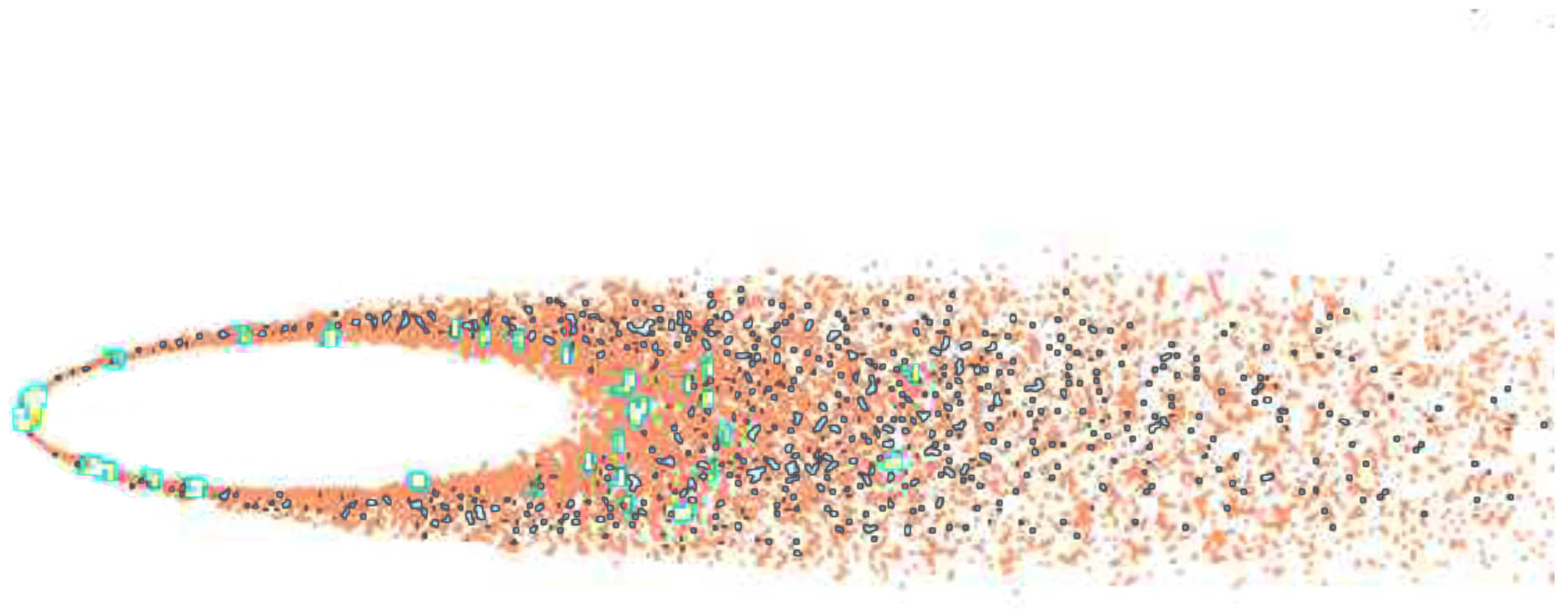}}
		\subfigure[$q=0.1 R_{\odot}$.]
		{\label{fig:collisions0_1q}\includegraphics[scale=0.475]{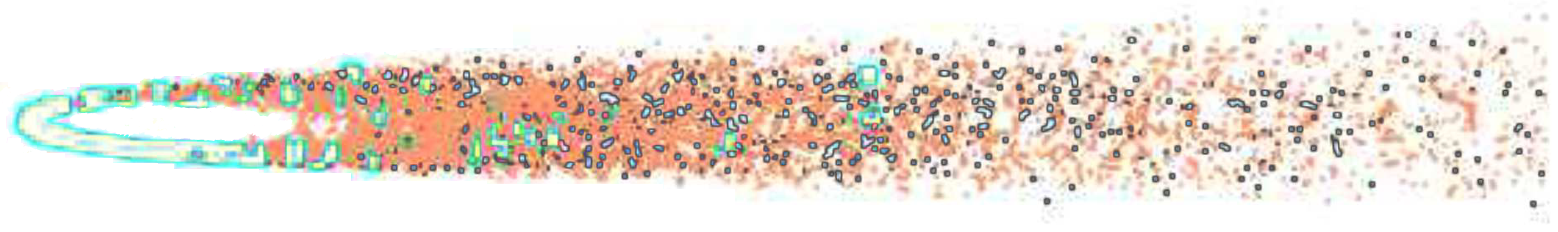}}	
	\end{center}
	\caption{Top down view, tracking collision locations over a three months period (equivalent to $\sim$10 original planet orbits) in fully formed discs of Mars-sized tidally disrupted planets around a 0.6$M_\odot$ WD. The discs are initially obtained via full SPH simulations with $a=0.1$ AU, and $q$ of (a) 1$R_{\odot}$; (b) 0.5$R_{\odot}$ and (c) 0.1$R_{\odot}$. Scale is shown in Panel (a). Collisions are depicted by blue circles. Other colours denote composition: orange - rock ; black - iron ; WD is omitted since obscured in this scale for panels (b) and (c).}
	\label{fig:collisions}
\end{figure}

The spatial distribution of collisions is captured in Figure \ref{fig:collisions}. We track all the collisions that have occurred over a period of three months, which according to Section \ref{SS:DiscFormation}, is comparable to the typical disc formation timescale (i.e., the time it took these specific discs to fully form). The collision locations are shown in red, superimposed on the discs. Colour denotes composition: orange - rock ; black - iron. Resolution is 10K particles.

Most collisions occur near or even inside the Roche sphere, whereas in the outer regions of the disc collisions are far more rare. Even if we neglect spatial distribution and count the total number of collisions over the entire domain, they amount to no more than 0.38-1.68\% of the number of particles. In other words, we argue that in the timescale relevant to the application of our hybrid model, the disc is collisionless at the 99\% level.

In fact, Figure \ref{fig:collisions} estimates the amount of collisions in a fully formed, SPH-simulated disc, whereas the hybrid model begins with a sparse disc and only completes with a fully formed disc. Thus, during the hybrid model, the disc should be even less collisional than the estimate in Figure \ref{fig:collisions}. 

In fact, if we were to use $k$ times more particles, and hence $k$ times less massive particles, the cross-section for collision (which goes like the projected geometric area) of a single particle (assuming similar densities) would be smaller by a factor of $k^{-2/3}$. The total cross section for collisions of all particles would therefore increase by the new number of particles, times their individual cross-section, i.e. $k\times k^{-2/3}=k^{1/3}$ and we could therefore expect even far higher resolution simulations, with even up to hundreds of times larger number of particles to be reasonably collisionsless.

We conclude that our assumption of a collisionless disc is acceptable, considering the relevant time duration in which our disc initially forms. We note however that in the longer-term evolution of the disc, in which it shrinks and circularizes, we should be mindful of the collisions near the star and their possible contribution. For example, \cite{KenyonBromley-2017,KenyonBromley-2018} show that when the disc reaches a much more compact state (in which $e \sim 0.01$), collisional grinding quickly reduces the particles in the disc to mere dust and gas. When our simulations conclude, the bound disc is still very eccentric ($e>0.9$), but slow grinding of particles may, on longer timescales (see also discussion in Section \ref{SS:CircularizationEvolution}), strengthen the influence of PR drag, sublimation, etc.

\end{appendices}

\bsp	
\label{lastpage}
\end{document}